\documentclass[submitting]{nst}
\usepackage{subfigure,dcolumn}
\usepackage[T2A,T1]{fontenc}

\usepackage{epstopdf}
\usepackage{mhchem}
\usepackage{upgreek}
\usepackage{tabularx}
\usepackage{bm}
\usepackage{amsmath} \usepackage{braket} \usepackage{epsfig}
\usepackage{tensor}
\usepackage{CJKutf8}
\usepackage{array}
\usepackage{titlesec}
\usepackage{floatrow}
\usepackage{ifthen}
\usepackage{sidecap}
\usepackage[para,online,flushleft]{threeparttablex}

\usepackage{xr-hyper}
\usepackage{hyperref}
\hypersetup{breaklinks=true,colorlinks=true,linkcolor=blue,citecolor=blue,filecolor=magenta,urlcolor=cyan}


\usepackage{filecontents}
\usepackage[all]{hypcap}
\usepackage{graphicx}
\usepackage{orcidlink}

\usepackage{xcolor}
\definecolor{pastelgray}{rgb}{0.81, 0.81, 0.77}
\definecolor{beaublue}{rgb}{0.9, 0.9, 0.93}
\definecolor{lrpcyan}{RGB}{208,247,244}
\definecolor{lrpblue}{RGB}{0,112,160}
\definecolor{lrporange}{RGB}{246,166,72}
\definecolor{lrpgreen}{RGB}{225,242,218}
\definecolor{lrpgray}{RGB}{245,245,242}
\definecolor{lrpred}{RGB}{176,62,47}

\newsavebox{\facilitysidebarcontent}
\newsavebox{\planningboxcontent}
\newcommand{\facilitytag}[1]{%
    \setlength{\fboxsep}{3pt}%
    \noindent\fcolorbox{black}{lrporange}{\textbf{#1}}%
}
\newcommand{\facilityentry}[3]{%
    \facilitytag{#1}\par\vspace{0.25em}
    \textbf{#2}\par
    #3\par\vspace{0.75em}
}
\newcommand{\facilityimage}[2]{%
    \begin{center}
    \includegraphics[width=0.95\linewidth]{#1}\par
    {\scriptsize #2}
    \end{center}
    \vspace{0.35em}
}
\newenvironment{facilitysidebar}[1]{%
    \begin{center}
    \setlength{\fboxsep}{7pt}%
    \begin{lrbox}{\facilitysidebarcontent}%
    \begin{minipage}{0.93\columnwidth}
    \small\raggedright\setlength{\parindent}{0pt}
    {\bfseries\color{lrpblue}#1}\par\vspace{0.45em}
}{%
    \end{minipage}%
    \end{lrbox}%
    \noindent\colorbox{lrpcyan}{\usebox{\facilitysidebarcontent}}%
    \end{center}
}
\newenvironment{recommendationbox}[1]{%
    \begin{center}
    \setlength{\fboxsep}{7pt}%
    \begin{lrbox}{\planningboxcontent}%
    \begin{minipage}{0.93\columnwidth}
    \small\raggedright\setlength{\parindent}{0pt}
    {\bfseries\color{lrpred}#1}\par\vspace{0.35em}
}{%
    \end{minipage}%
    \end{lrbox}%
    \noindent\colorbox{lrpgreen}{\usebox{\planningboxcontent}}%
    \end{center}
}

\usepackage{listings}
\usepackage{tikz,xcolor,hyperref}

\definecolor{lime}{HTML}{A6CE39}
\DeclareRobustCommand{\orcidicon}{
	\begin{tikzpicture}
		\draw[lime, fill=lime] (0,0) 
		circle [radius=0.16] 
		node[white] {{\fontfamily{qag}\selectfont \tiny ID}};
		\draw[white, fill=white] (-0.0625,0.095) 
		circle [radius=0.007];
	\end{tikzpicture}
	\hspace{-2mm}
}
\foreach \x in {A, ..., Z}{
	\expandafter\xdef\csname orcid\x\endcsname{\noexpand\href{https://orcid.org/\csname orcidauthor\x\endcsname}{\noexpand\orcidicon}}
}

\begin{document}
\makeatletter
\let\switch@array\relax
\makeatother

\begin{CJK*}{UTF8}{gbsn}
    
\title{Frontier Questions and Emerging Directions in Nuclear Science and Technology}

\author{Y. G. Ma~\orcidlink{0000-0002-0233-9900}}\email{mayugang@fudan.edu.cn}
\affiliation{School of Physics, East China Normal University, Shanghai 200241, China}
\affiliation{Key Laboratory of Nuclear Physics and Ion-beam Application (MOE), Institute of Modern Physics, Fudan University, Shanghai 200433, China}
\affiliation{Shanghai Research Center for Theoretical Nuclear Physics, NSFC and Fudan University, Shanghai 200438, China}

\begin{abstract}
Recent advances in nuclear science and technology are being driven simultaneously by fundamental questions on strong interactions and many-body emergence, by the rapid expansion of rare-isotope capabilities and multimessenger astronomy, and by growing societal demand for clean energy, precision medicine, and strategic technologies. This review reorganizes the “ten frontier questions” for nuclear science and technology, offering a scholarly roadmap accessible to a broad audience. We first discuss the fundamental frontiers, including the nonperturbative origin of hadronic mass, the properties of QCD matter under extreme conditions, the multiscale evolution of nuclear structure from light nuclei to the superheavy region, the physics of exotic nuclei and open quantum systems near the driplines, and the nuclear-astrophysical origin of the elements. We then emphasize enabling methodologies, especially modern \textit{ab initio} and continuum-coupled theories, advanced accelerator and detector platforms, precision mass spectrometry, and the emerging role of data-driven and artificial-intelligence-assisted methodologies. Finally, we review translational and strategic directions, including advanced fission and fusion energy systems, cross-disciplinary nuclear technologies such as radiomedicine, isotope science, and nuclear clocks, as well as the long-term challenges of fuel cycles, waste management, and international cooperation. Rather than serving as an exhaustive bibliography of each subfield, this article aims to provide an integrative research framework that connects frontier scientific problems with enabling infrastructure, application scenarios, and long-range strategic planning.
\end{abstract}


\date{\today}
\maketitle

\onecolumngrid
\setcounter{tocdepth}{2}
\tableofcontents
\twocolumngrid

\section{Introduction}

\begin{table*}[t]
\centering
\caption{Facility-centered implementation matrix for a China long-range plan.}
\label{tab:china-lrp-matrix}

\begingroup
\small
\setlength{\tabcolsep}{3pt}
\renewcommand{\arraystretch}{1.15}
\newcommand{\Tcell}[2]{%
  \parbox[t]{#1}{\raggedright #2\par}%
}

\begin{tabular}{@{}cccc@{}}
\hline
\hline
\Tcell{0.18\textwidth}{\textbf{Strategic area}} &
\Tcell{0.21\textwidth}{\textbf{Domestic facility base}} &
\Tcell{0.28\textwidth}{\textbf{Science and technology deliverables}} &
\Tcell{0.24\textwidth}{\textbf{Planning priority}} \\
\hline

\Tcell{0.18\textwidth}{QCD matter and dense nuclear matter} &
\Tcell{0.21\textwidth}{HIAF, HIRFL-CSR, heavy-ion detector systems, theory centers} &
\Tcell{0.28\textwidth}{Equation of state, symmetry energy, collective flow, strangeness, hypernuclei, rare probes, links to compact stars} &
\Tcell{0.24\textwidth}{High-rate detectors, calibrated transport codes, common analysis challenges, open benchmark data} \\

\Tcell{0.18\textwidth}{Rare isotopes and nuclear structure} &
\Tcell{0.21\textwidth}{HIAF, HIRFL-CSR, storage rings, recoil separators, traps, decay arrays} &
\Tcell{0.28\textwidth}{Dripline nuclei, shell evolution, halos, superheavy and neutron-rich heavy nuclei, open quantum systems} &
\Tcell{0.24\textwidth}{Beam-time concentration, next-generation separators, active targets, gamma and neutron arrays} \\

\Tcell{0.18\textwidth}{Nuclear astrophysics} &
\Tcell{0.21\textwidth}{CJPL/JUNA, HIAF storage rings, SLEGS, CSNS/Back-n} &
\Tcell{0.28\textwidth}{Stellar burning rates, breakout reactions, p-process and r-process constraints, neutron-capture data} &
\Tcell{0.24\textwidth}{Prioritized reaction list, evaluated rates, uncertainty propagation to stellar and multimessenger observables} \\

\Tcell{0.18\textwidth}{Photonuclear and neutron data} &
\Tcell{0.21\textwidth}{SSRF/SLEGS, CSNS/Back-n, gamma and neutron detector arrays} &
\Tcell{0.28\textwidth}{Photonuclear cross sections, fission and capture data, isotope production, safeguards and reactor data} &
\Tcell{0.24\textwidth}{National evaluated-data pipeline with metadata, covariance, detector response, and reproducible workflows} \\

\Tcell{0.18\textwidth}{Advanced nuclear energy} &
\Tcell{0.21\textwidth}{TMSR, CIADS, EAST/CFETR-related fusion programs, laser fusion, materials platforms} &
\Tcell{0.28\textwidth}{Thorium cycle, molten salts, ADS transmutation, fusion nuclear science, irradiation materials, waste reduction} &
\Tcell{0.24\textwidth}{Integrated fuel-cycle, materials, tritium, safety, and digital-control roadmap} \\

\Tcell{0.18\textwidth}{Isotopes, medicine, and precision} &
\Tcell{0.21\textwidth}{Reactor and accelerator isotope sources, radiochemistry labs, nuclear clocks, microcalorimeters} &
\Tcell{0.28\textwidth}{Radiopharmaceuticals, theranostics, isotope metrology, nuclear clocks, environmental tracing} &
\Tcell{0.24\textwidth}{Stable isotope supply chains, clinical translation, metrology standards, precision spectroscopy platforms} \\

\hline
\hline
\end{tabular}
\endgroup
\end{table*}

Nuclear science and technology occupies a uniquely broad and interconnected position within modern physical science. It lies at the intersection of fundamental quantum many-body physics, large-scale experimental infrastructure, and strategically important technological applications. On the one hand, it addresses some of the deepest questions in contemporary physics, including the origin of visible mass from quantum chromodynamics~\cite{Ding,Lorce}, the emergence of complex structure in strongly interacting finite systems~\cite{Ye2023,Freer2007RMP,Holt2013JPhysG}, the limits of nuclear existence~\cite{Erler2012NuclearLandscape,PhysRevLett.133.222501,PhysRevLett.135.012501,Charity2023}, and the behavior of nuclear matter under extreme conditions~\cite{ChenJH2024,Shou2024,PPNP2,Multi,WangR}. On the other hand, it provides the scientific foundation for a wide range of societal and technological domains, including nuclear energy systems~\cite{Abram2008GenerationIV,TMSR,Hesch2024FusionProgress}, isotope production for medicine and industry~\cite{SLEGS-iso,SLEGS2,Kolos2022NuclearDataNeeds}, precision timekeeping and metrology~\cite{RN115,Yamaguchi2024,Berengut2025IsotopeShift,Pomme2022Radionuclide}, radiological diagnostics and therapy~\cite{RMP,Zhang2025Radiopharmaceuticals,Sgouros2020RPT}, national security technologies~\cite{Kolos2022NuclearDataNeeds,AlHamrashdi2019NuclearSecurity}, and advanced imaging techniques~\cite{Wu2024fMetaTPC,Beceiro2015,Miernik2007b,Blank2010}.

This dual identity—simultaneously fundamental and applied—has long been a defining feature of the field~\cite{nsac2023lrp,nupecc2024lrp,Ma2025Hotspot,Ma_Book,Nupeec}. However, recent decades of progress have significantly sharpened the need for a more integrated conceptual framework. Traditional divisions among nuclear structure, nuclear reactions, nuclear astrophysics, high-energy nuclear matter, and applied nuclear technologies are increasingly inadequate to describe the interconnected nature of modern research~\cite{Ma2025Hotspot,Navratil2016,Xu2024}. Advances in experimental capability, theoretical modeling, and computational power have revealed that many of the most important scientific challenges cut across these boundaries~\cite{Zhou2022,Ekstrom2023AbInitio,Boehnlein2022}. As a result, nuclear science is progressively evolving toward an integrated, systems-level perspective in which foundational physics, enabling technologies, and application-driven questions are treated within a coherent intellectual structure~\cite{nsac2023lrp,nupecc2024lrp,Ma2025Hotspot}.

A particularly useful organizing principle for this evolving landscape is provided by the recently proposed ``ten frontier questions'' in nuclear science and technology \cite{Ma2025}. While originally formulated as a broad conceptual guide, these questions in fact delineate a structured and internally consistent research agenda that reflects the current state and future trajectory of the field. They naturally align with international long-range planning efforts in nuclear physics, as reflected in major roadmap documents from global communities \cite{nsac2023lrp,nupecc2024lrp}. At the same time, they resonate strongly with recent topical reviews and hotspot analyses that highlight rapid progress in areas such as rare-isotope beam physics, relativistic heavy-ion collisions, nuclear astrophysics, precision mass measurements, nuclear energy systems, and advanced accelerator technologies \cite{Ma2022Hotspot,Ma2023Hotspot,Ma2024Hotspot,Ma2025Hotspot,Ye2025,ChenJH2024,Shou2024,LiuNST2024}.

The present article is intended as a perspective synthesis rather than a conventional topical review. Our goal is not to present the ten frontier questions in a descriptive or popular format, but to reinterpret them in terms of a unified research framework suitable for a professional scientific audience. In doing so, we aim to identify the key physical drivers that underlie these questions, to clarify their interconnections, and to highlight the methodological and infrastructural developments that are enabling progress across multiple subfields.

\begin{figure*}[htb]
\floatbox[{\capbeside\thisfloatsetup{capbesideposition={right,top},capbesidewidth=0.2\textwidth}}]{figure}[\FBwidth]
{\caption{Schematic organization of the ten frontier nuclear questions. The three overlapping primary-color circles represent the three main objectives of the discussion: scientific motivations, enabling methodologies, and strategic and societal coupling. The central overlap highlights the shared core of the framework, ``Ten Frontier Nuclear Questions.'' Numbered icons summarize the major topics, including QCD and nucleon mass generation, extreme nuclear matter, nuclear structure and reaction dynamics, exotic nuclei, cosmic nucleosynthesis, fission and fusion energy, nuclear techniques and crossovers, fuel cycle and waste management, next-generation facilities, and international cooperation and governance. The overlap regions emphasize that these questions are interconnected through common theoretical, experimental, and technological threads. Original schematic prepared by the author.}\label{fig:frontier_nuclear_question}}
{\includegraphics[width=0.75\textwidth]{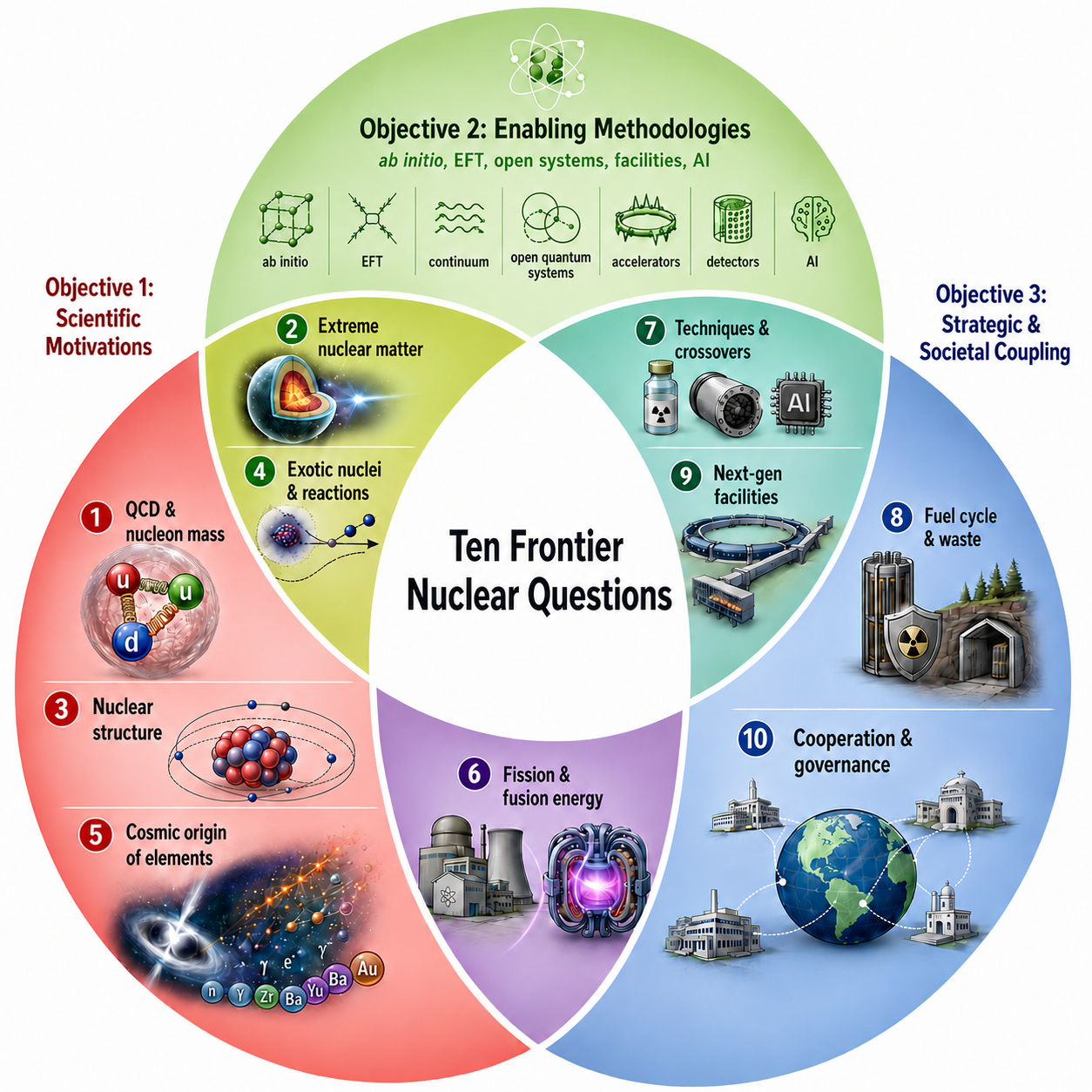}}
\end{figure*}

As shown in Fig.~\ref{fig:frontier_nuclear_question}, we structure the discussion around three main objectives. First, we analyze the fundamental scientific motivations behind each frontier question and situate them within the broader context of modern nuclear physics. This includes the role of quantum chromodynamics in shaping nuclear structure, the emergence of collective behavior in finite quantum systems, the properties of nuclear matter under extreme temperature and density, and the astrophysical origin of the elements. Second, we examine the enabling methodologies that increasingly cut across traditional subfield boundaries. These include \textit{ab initio} many-body theory, effective field theory approaches, continuum-coupled and open quantum system descriptions, large-scale accelerator facilities, advanced detector systems, precision metrology techniques, and data-driven and AI-assisted computational tools. Third, we discuss the growing coupling between fundamental nuclear science and broader strategic and societal applications, particularly in areas such as sustainable nuclear energy, nuclear medicine, isotope technology, and long-term nuclear governance.

{This Perspective is not an exhaustive review of each subfield, but a selective and representative roadmap connecting frontier questions, facilities, technologies, and strategic needs.} Each of the major subfields discussed here has developed into a mature discipline with its own extensive literature, specialized methodologies, and detailed technical reviews. A comprehensive treatment of all aspects would therefore be beyond the scope of a single article. Instead, we aim to provide a structured “research map” that highlights the most relevant connections between foundational scientific questions, representative recent advances, and key unresolved challenges.

In this sense, the ten frontier questions serve not only as a thematic organization of current nuclear science, but also as a conceptual bridge linking microscopic theory, experimental innovation, computational methods, and societal applications. By examining these questions within a unified framework, we hope to clarify how nuclear science is evolving from a collection of partially separated subdisciplines into an increasingly coherent and integrated field of research, in which progress in one domain often drives and {stimulates} progress in others. {Throughout the article, the numbered sidebars provide a second reading layer: they translate the thematic discussion into facility capabilities, experimental deliverables, and planning dependencies without interrupting the main scientific narrative.}

\section{The ten frontier questions as an organizing framework}
The ten frontier questions may be grouped into six basic-science themes and four application- or strategy-oriented themes (see Fig.~\ref{fig:frontier_nuclear_question}). In condensed form, they ask:
\begin{enumerate}
    \item How does QCD generate nucleon mass, and how do confinement and chiral symmetry breaking manifest across scales?
    \item {What are the phases and transport properties of nuclear matter under extreme temperature and baryon density, in the strong electromagnetic fields and vorticity generated in heavy-ion collisions, and in the gravitational and magnetic environments of compact stars?}
    \item How do shell evolution, clustering, deformation, collectivity, and superheavy stability emerge within a unified description of nuclear structure?
    \item What new phenomena and reaction mechanisms appear in exotic nuclei far from the valley of $\beta$ stability?
    \item How do astrophysical environments and weak interactions shape the cosmic origin of the elements?
    \item How can advanced fission and controlled fusion contribute to a safe, clean, and sustainable energy system?
    \item How can nuclear techniques further integrate with medicine, materials, precision measurement, quantum technologies, and artificial intelligence?
    \item Can spent fuel management, waste disposal, and emergency-response systems be organized into a long-term and system-level solution?
    \item How can next-generation accelerators, detectors, and intelligent experimental platforms reshape the research paradigm?
    \item How should international cooperation and strategic competition be balanced in large facilities, frontier experiments, and nuclear governance?
\end{enumerate}

These questions are not independent. They are connected by several recurring threads: nonperturbative quantum many-body dynamics, extreme conditions and open systems, precision observables enabled by large facilities, and a strong feedback loop between fundamental discovery and technological application (see Fig.~\ref{fig:frontier_nuclear_question}).

{Table~\ref{tab:china-lrp-matrix} summarizes how representative Chinese facilities connect the ten frontier questions to measurable science deliverables and long-range planning priorities. Relative to facility lists alone, this matrix makes the cross-cutting dependencies---shared detectors, nuclear data, theory, materials, and digital infrastructure---explicit. Here and throughout, HIAF denotes the High-Intensity Heavy-Ion Accelerator Facility.}

In what follows, we discuss these questions in a more integrated way, beginning with the basic-science frontiers.

\section{Fundamental frontiers: from QCD to the limits of nuclear existence}

\subsection{Strong interaction, confinement, and the origin of visible mass}

\begin{figure}[htbp]
\includegraphics[width=1.0\linewidth]{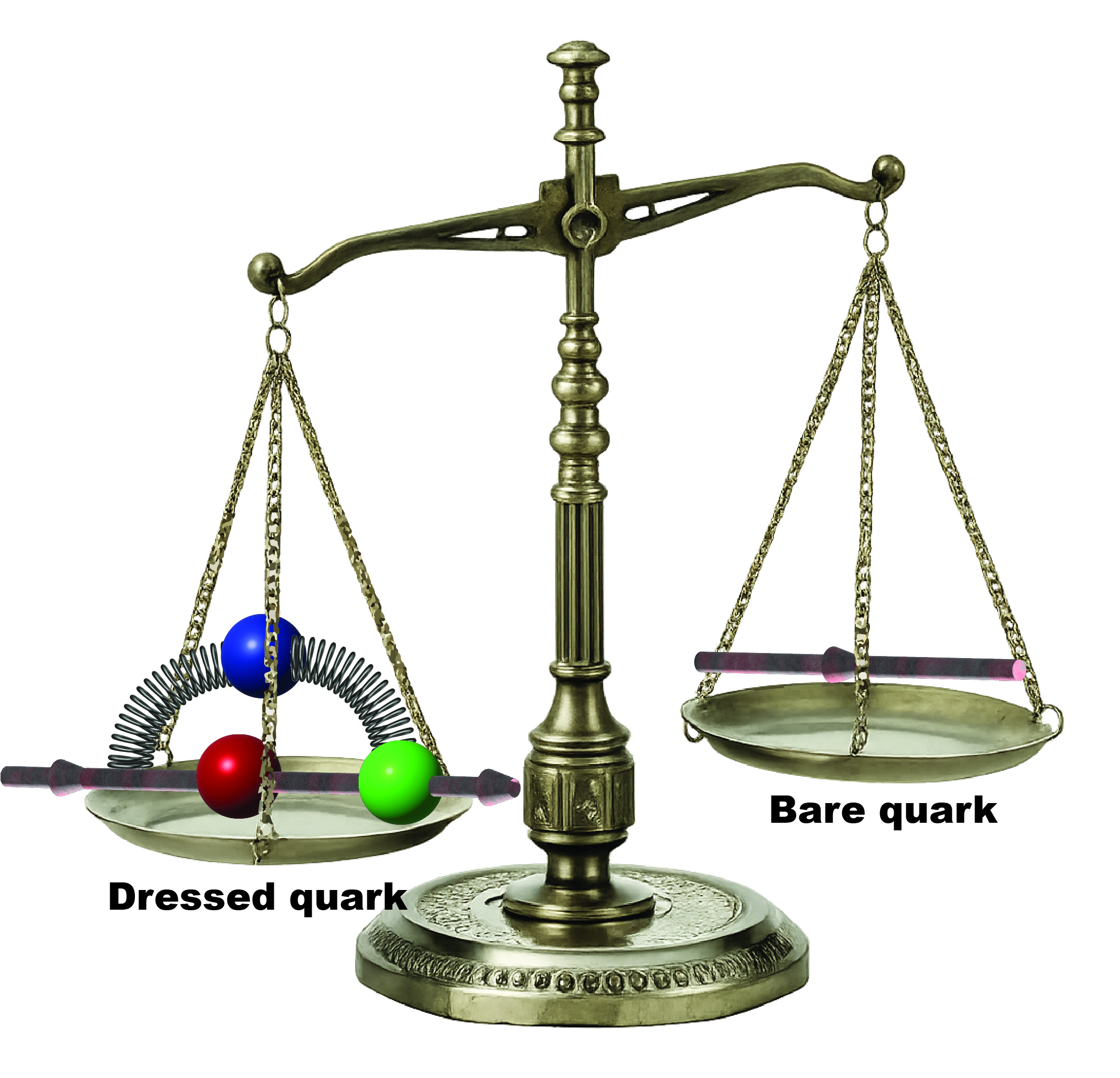}
\caption{\label{fig:strong_interaction}
Schematic illustration of how the strong interaction generates the effective mass of quarks. At short distances, corresponding to the perturbative regime of QCD, the quark behaves approximately as a bare quark with a small current mass. As the distance scale increases toward hadronic scales, nonperturbative QCD effects become dominant, and the quark is progressively dressed by gluon fields and quark–antiquark excitations, leading to a much larger effective (constituent) mass. The balance emphasizes the strong enhancement of quark mass through confinement-related dynamics and dynamical chiral symmetry breaking, which together underlie the origin of most of the visible mass in the universe.
{Schematic redrawn and adapted by the author from a Jefferson Lab illustration by Shannon West~\cite{Phys_org_news}.}
}
\end{figure}

A central ambition of modern nuclear physics is to understand how the strong interaction, as described by quantum chromodynamics (QCD), generates the observed properties of hadrons and nuclei from its fundamental quark and gluon degrees of freedom. Among these properties, the origin of the mass of ordinary visible matter stands out as one of the most profound and conceptually rich problems~\cite{Ding} (Fig.~\ref{fig:strong_interaction}). It is now well established that the masses of protons and neutrons—the building blocks of atomic nuclei—are not primarily determined by the small current masses of the up and down quarks, which contribute only a few percent to the total nucleon mass. Instead, the dominant contribution arises from the complex, nonperturbative dynamics of QCD, including gluon self-interactions, quark kinetic and potential energy, and the phenomenon of spontaneous chiral symmetry breaking~\cite{Ding,Lorce}. Understanding how these elements combine to generate mass is not only a question of quantitative decomposition, but also of identifying the underlying mechanisms and emergent structures that give rise to the bulk of visible matter in the universe.

This problem lies at the intersection of several major areas of nuclear and particle physics, including hadron structure, hadron spectroscopy, and the theory of nuclear forces \cite{ZhuPR,GuoRMP,ChenHadron,Machleidt2024ChiralEFT}. It reflects the broader concept of emergence in strongly interacting systems, where simple underlying laws lead to rich and often unexpected macroscopic behavior. In QCD, the emergence of hadrons as color-neutral bound states, the formation of mass gaps, and the appearance of effective degrees of freedom at low energies are all manifestations of this principle. Consequently, the study of mass generation is deeply intertwined with understanding how confinement—the absence of free quarks and gluons—and dynamical chiral symmetry breaking operate in tandem within the QCD vacuum~\cite{Ding,Lorce}.

A key challenge in addressing these questions is the inherently multiscale nature of the strong interaction. At very short distances or high momentum transfers, QCD exhibits asymptotic freedom, allowing perturbative techniques to be applied successfully. In this regime, quarks and gluons behave as weakly interacting particles, and many observables can be calculated using systematic expansions in the strong coupling constant. However, as one moves to larger distance scales characteristic of hadrons (on the order of a femtometer), the coupling becomes strong, and perturbative methods break down. It is in this nonperturbative regime that confinement and dynamical chiral symmetry breaking emerge, giving rise to the rich spectrum of hadrons and their internal structure.

Importantly, confinement and chiral symmetry breaking should not be viewed as isolated or independent mechanisms, but rather as collective phenomena that manifest themselves across a wide range of observables. For example, hadron mass spectra encode information about the underlying QCD dynamics through patterns of excitation and symmetry breaking. Elastic and transition form factors provide insight into the spatial distribution of charge and magnetization within hadrons, while generalized parton distributions (GPDs) and transverse momentum distributions (TMDs) offer a multidimensional picture of quark and gluon structure in momentum and coordinate space. Together, these observables form a comprehensive framework for probing the internal dynamics of hadrons and the mechanisms responsible for mass generation. {A practical near-term frontier is to determine how quarks and gluons share the nucleon's momentum and spin through one-dimensional parton distributions, three-dimensional TMD and GPD imaging, and ultimately phase-space descriptions such as Wigner distributions. This hierarchy of one-, three-, and five-dimensional information is a central motivation for the new generation of electron--ion colliders, including the EIC and the proposed EicC, and connects the ultimate mass-origin question to experimentally accessible tomography of sea quarks and gluons~\cite{Anderle2021,Accardi2016EIC}.}

From the perspective of nuclear science, the problem extends beyond individual hadrons to the emergence of nuclear forces and many-body interactions. While QCD is the fundamental theory, nuclei are effectively described in terms of nucleons interacting via forces that reflect underlying quark-gluon dynamics \cite{Machleidt2024ChiralEFT}. One of the central goals of modern nuclear theory is therefore to derive nucleon–nucleon and many-nucleon interactions from QCD in a systematically improvable way. Approaches such as chiral effective field theory (EFT) provide a powerful framework for connecting QCD symmetries to low-energy nuclear interactions, enabling controlled expansions and uncertainty quantification \cite{Machleidt2024ChiralEFT}. At the same time, lattice QCD calculations are beginning to provide direct insights into hadron masses, interactions, and even light nuclear systems from first principles, offering a bridge between fundamental theory and nuclear phenomenology.

Despite significant progress, many aspects of the mass generation problem remain open \cite{Reinosa,Lukashov}. For instance, a complete and universally accepted decomposition of the nucleon mass in terms of quark and gluon contributions is still under active investigation. The role of gluon dynamics, in particular, is both dominant and subtle, as gluons contribute not only through their energy density but also through quantum effects such as the trace anomaly of the QCD energy-momentum tensor. Similarly, the interplay between confinement and chiral symmetry breaking—whether one drives the other or whether they emerge simultaneously from a deeper mechanism—remains a subject of ongoing theoretical debate.

Meanwhile, studies of relativistic heavy-ion collisions, for example, provide valuable constraints on the properties of QCD matter under extreme conditions, including temperature, density, and vorticity. Recent assessments of QCD matter and spin-related observables in such collisions illustrate how the field is moving toward precision measurements that can probe the nonperturbative regime of the strong interaction \cite{ChenJH2024,Shou2024,Liang1,Liang4,Liang5}. Observables such as global polarization, spin alignment, and fluctuations of conserved charges offer new windows into the role of quark-gluon dynamics and symmetry breaking in strongly interacting matter.

Furthermore, these developments highlight an important conceptual continuity: the same underlying QCD dynamics that generate mass in the vacuum also govern the behavior of matter under extreme conditions. In this sense, the study of hot and dense QCD matter complements and enriches our understanding of hadron structure, rather than constituting a separate line of inquiry. This perspective is increasingly reflected in long-range planning documents and community roadmaps, where hadron structure, QCD matter, and emergent nuclear phenomena are treated as tightly connected pillars of a unified research program \cite{nsac2023lrp,nupecc2024lrp}. Such integrative approaches emphasize the importance of combining experimental, theoretical, and computational advances to address the most fundamental questions in nuclear physics.

In this broader context, the question of the origin of visible mass can be viewed as a “gateway problem” that encapsulates many of the key challenges and opportunities in the field. It motivates the development of new theoretical tools, such as advanced lattice QCD techniques, continuum methods like Dyson–Schwinger equations, and improved effective field theories. At the same time, it drives the design of next-generation experimental facilities, including electron-ion colliders and upgraded hadron spectroscopy programs, which aim to provide high-precision data on hadron structure and dynamics.

Ultimately, solving the mass origin problem will require a coordinated effort across multiple frontiers, combining insights from theory, experiment, and computation. It will also demand a deeper conceptual understanding of how complex phenomena emerge from the fundamental laws of QCD. As such, it represents not only a central challenge for nuclear physics, but also a cornerstone in our broader quest to understand the structure and evolution of the visible universe.

\subsection{Nuclear matter under extreme conditions and the QCD phase diagram}

\begin{figure}[htbp]
\includegraphics[width=0.94\linewidth]{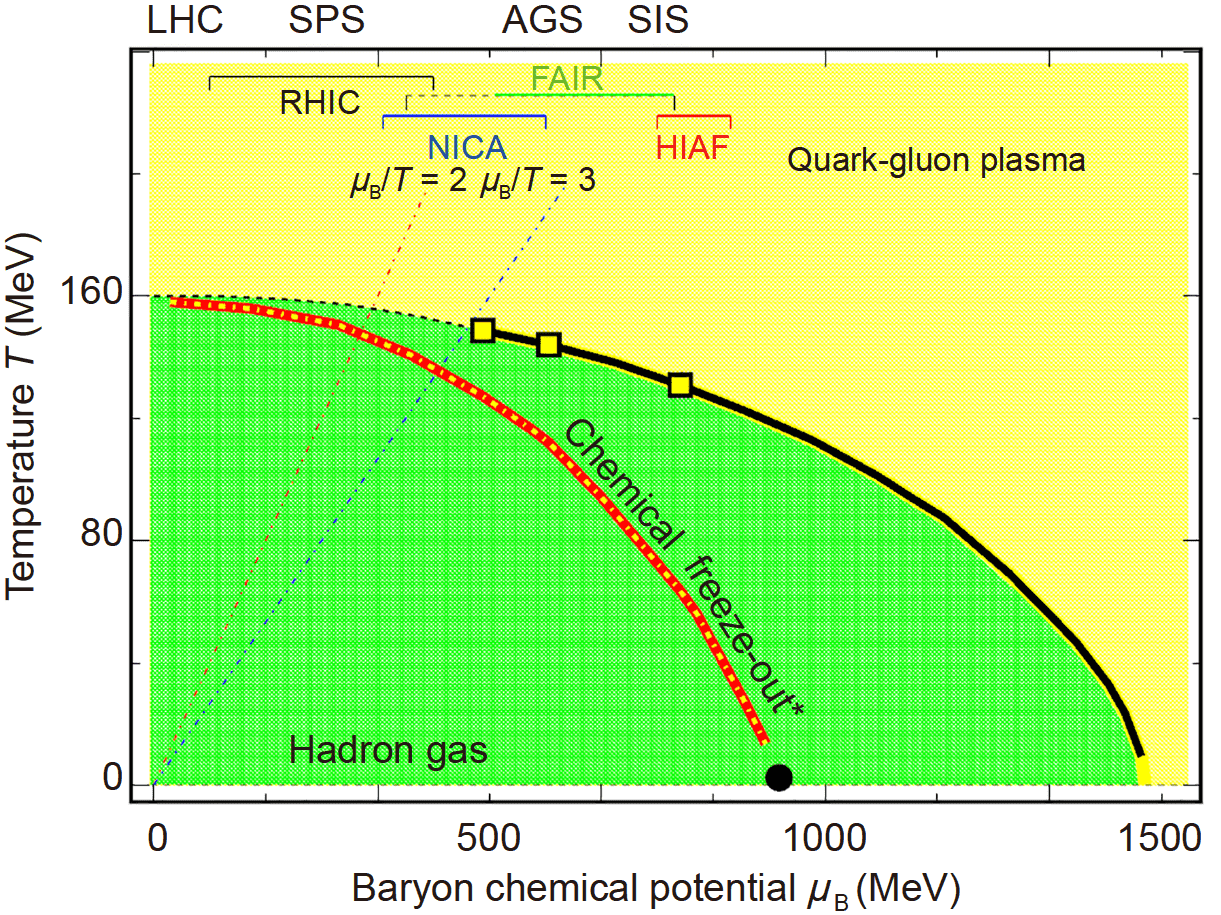}
\caption{\label{fig:QCD_phase}
{Schematic diagram of the QCD phase diagram \cite{Ma2020}. Vertical- and horizontal-axis are the temperature $T$, in MeV and baryon chemical potential $\mu_{\rm B}$, in MeV. The solid and dashed black lines represent the phase boundary and smooth crossover from QGP to hadron gas, respectively. The square point is the possible critical point. The empirical results of chemical freeze-out are shown as the red-line. Heavy-ion collision facilities of LHC (Europe), RHIC (USA), NICA (Russia), FAIR (Germany) and HIAF (China) with corresponding covered regions of chemical potential are also shown at the top of the figure.}
}
\end{figure}

The study of nuclear matter under extreme conditions seeks to understand how strongly interacting matter behaves when subjected to environments far beyond those encountered in ordinary nuclei \cite{PBM,HuangM,ChenJH2024,Shou2024}. These conditions include extremely high temperatures, such as those that existed microseconds after the Big Bang; very high baryon densities, characteristic of the interior of neutron stars; and intense angular momentum and electromagnetic fields, which can be transiently generated in non-central relativistic heavy-ion collisions~\cite{Bzdak2019PhysRept,Multi,Liang1,STAR2,Liang4}. Exploring these regimes provides a unique opportunity to probe the fundamental properties of quantum chromodynamics (QCD), particularly in its nonperturbative domain, where the interplay between quarks and gluons gives rise to collective phenomena such as deconfinement, chiral-symmetry restoration, and emergent hydrodynamic behavior~\cite{PBM,HuangM,ChenJH2024,Shou2024}.

Relativistic heavy-ion collisions serve as the primary experimental tool for recreating such extreme conditions in the laboratory. Facilities such as the Relativistic Heavy Ion Collider (RHIC) and the Large Hadron Collider (LHC) enable collisions of heavy nuclei at ultrarelativistic energies, producing short-lived fireballs of deconfined quark-gluon matter~\cite{ChenJH2024,Shou2024}. These fireballs reach temperatures exceeding several hundred MeV, well above the predicted crossover temperature for the transition between hadronic matter and the quark-gluon plasma (QGP) predicted by lattice QCD~\cite{Philipsen,HuangM}. At lower collision energies, experiments can access regions of higher baryon density, allowing for systematic exploration of different regions of the QCD phase diagram~\cite{Bzdak2019PhysRept,ChenJH2024}. Although the lifetime of the produced matter is extremely short (on the order of 10 fm/$c$), a wealth of observables can be measured, providing insight into both equilibrium properties and dynamical evolution~\cite{ChenJH2024,Shou2024}.

The resulting research program spans a broad range of topics, including the mapping of the QCD phase diagram  (see Fig.~\ref{fig:QCD_phase}), the determination of transport coefficients such as shear and bulk viscosity, the study of hadronization mechanisms, and the investigation of how microscopic interactions lead to macroscopic collective behavior~\cite{Bzdak2019PhysRept,PBM,HuangM,ChenJH2024,Shou2024}. One of the most striking discoveries in this field is the near-perfect fluidity of the quark-gluon plasma, which exhibits extremely low shear viscosity relative to entropy density~\cite{HeinzSnellings2013FlowViscosity}. This behavior challenges early expectations of a weakly interacting quark-gluon gas and instead points to a strongly coupled medium with emergent hydrodynamic properties. Understanding the origin of this collective behavior and its connection to underlying QCD dynamics remains a central objective.

Recent experimental programs, particularly the Beam Energy Scan (BES) at RHIC and the precision measurements carried out by the ALICE collaboration at the LHC, have highlighted several frontier issues. These include the search for critical phenomena associated with a possible QCD critical point, the identification of the onset of deconfinement as collision energy increases, and the precise determination of thermodynamic properties such as temperature, pressure, and viscosity of the produced matter \cite{Stephanov1998PRL,Bzdak2019PhysRept,LuoXu2017NST,ChenJH2024,Shou2024}. Event-by-event fluctuations of conserved quantities, such as net baryon number, electric charge, and strangeness, are of particular interest, as they may carry signatures of critical behavior and phase transitions. However, disentangling such signals from background effects and finite-size dynamics remains a significant experimental and theoretical challenge. {A related unresolved issue is whether high-multiplicity proton-proton and proton-nucleus collisions create short-lived QGP droplets or whether their collective-like signatures can be explained by initial-state correlations and few-body final-state dynamics. System-size scans and observables with controlled nonflow sensitivity are therefore essential for identifying the minimum conditions for hydrodynamic behavior.}

Another rapidly developing frontier concerns the role of spin degrees of freedom and vorticity in heavy-ion collisions. Non-central collisions generate extremely large angular momentum, a fraction of which is transferred to the produced medium, leading to fluid vorticity and spin polarization effects. Observables such as global polarization of hyperons and spin alignment of vector mesons have evolved from qualitative discoveries into quantitative probes of the underlying dynamics. Measurements of global polarization indicate that the QGP may be the most vortical fluid ever observed, while deviations from naive expectations in vector-meson spin alignment suggest the presence of strong local fields and complex hadronization processes \cite{STAR1,STAR2,Liang1,Liang4,Becattini2013PRL,Becattini2021AnnRev,HuangXG2025}. These phenomena provide new avenues for studying the coupling between spin, orbital angular momentum, and the strong interaction, opening a novel dimension in the exploration of QCD matter \cite{STAR3,ShengXL}.

\begin{facilitysidebar}{Sidebar 1. Collider and hadron-structure facilities}
\facilityimage{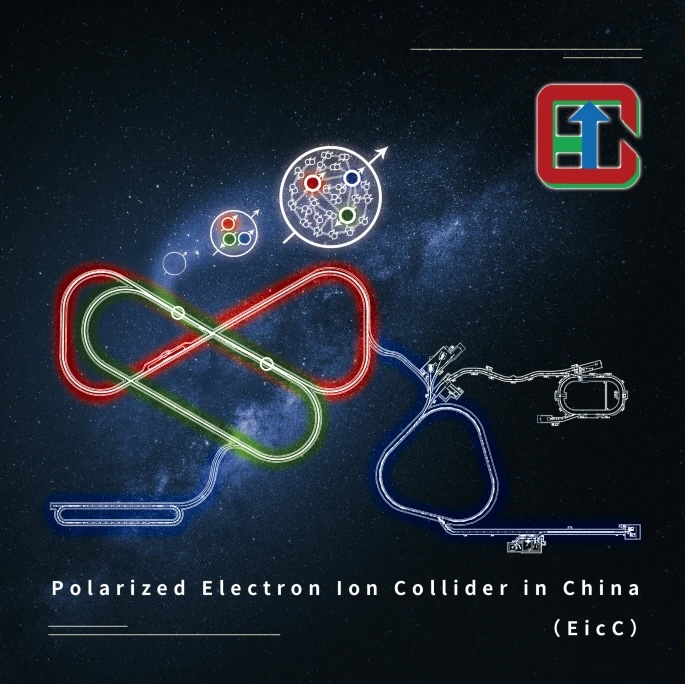}{A designed schematic plot for the Polarized electron-ion collider in China \cite{Anderle2021}.}
\facilityentry{Hadron mass and 3D structure}
{Electron-ion colliders and upgraded electron scattering programs}
{The central mission is to turn the qualitative question of mass generation into quantitative maps of quark and gluon structure. High-luminosity lepton--hadron scattering provides the cleanest route to generalized parton distributions, transverse-momentum distributions, gluon imaging, and the decomposition of nucleon mass and spin. The decisive observables include exclusive form factors, semi-inclusive hadron production, heavy-flavor channels, and energy-momentum-tensor constraints on pressure and shear distributions inside the nucleon. These measurements connect the mass-origin problem to the international EIC and upgraded fixed-target programs emphasized in long-range planning \cite{nsac2023lrp,nupecc2024lrp,Anderle2021}.}

\facilityentry{Hot and dense QCD matter}
{RHIC, LHC, and future heavy-ion programs}
{Relativistic heavy-ion facilities map the QCD phase diagram through anisotropic flow, event-by-event fluctuations, electromagnetic probes, heavy flavor, jets, spin polarization, and rare probes of the quark-gluon plasma \cite{ChenJH2024,Shou2024}. Their planning value lies in coordinated energy scans and detector upgrades: high baryon density requires flexible beam energy and high statistics, while early-time QGP dynamics require precision vertexing, calorimetry, and fast online reconstruction. Together with hadron spectroscopy and lattice-QCD programs, they form the facility class most directly tied to confinement, chiral symmetry breaking, and emergent QCD matter.}
\end{facilitysidebar}

In parallel, significant progress has been made in connecting heavy-ion collision observables to the properties of dense nuclear matter relevant for astrophysics. Transport-theory approaches, including Boltzmann-type models and relativistic hydrodynamics with hadronic afterburners, play a crucial role in bridging microscopic interactions and macroscopic observables. In particular, studies of symmetry-energy-sensitive observables in intermediate-energy heavy-ion collisions have improved constraints on the nuclear equation of state (EoS), especially its isospin-dependent component \cite{Deng,DingMQ2024,FangNT}. {Recent Bayesian and transport analyses further connect directed flow and FOPI observables to phase-transition scenarios, nuclear incompressibility, effective masses, and in-medium cross sections, while multimessenger inference tests the density dependence of covariant functionals~\cite{Wu2026PRD,Wei2026FOPI,Wei2026CDF}.} These constraints are essential for understanding the structure and evolution of neutron stars, including their mass-radius relation, cooling behavior, and response to mergers \cite{PPNP2,LiBA,WangR,WangJM,CaiBJ1,CaiBJ2}.

A major conceptual challenge in this field is the unification of the different regimes of QCD matter probed in laboratory experiments and astrophysical observations. Hot QCD matter created in heavy-ion collisions is characterized by high temperature and relatively low baryon density, while the matter inside neutron stars is cold but extremely dense. Despite these differences, both systems are governed by the same underlying theory and probe the response of strongly interacting matter far from normal nuclear conditions. Bridging these regimes requires a consistent description of the QCD equation of state over a wide range of temperature and density, as well as a deeper understanding of phase transitions, such as the possible transition to color-superconducting phases at high density \cite{Philipsen}.

Progress toward this goal will depend on advances across multiple fronts. Experimentally, higher-statistics measurements and new observables, including electromagnetic probes such as direct photons and dileptons, will provide more direct access to the early stages of the collision and the properties of the medium. Event-by-event analyses will continue to play a key role in extracting information about fluctuations and correlations, while improved detector capabilities will enhance sensitivity to rare processes and spin-related observables. Theoretically, developments in transport modeling, lattice QCD at finite baryon density, and effective field theories will be essential for interpreting experimental data and making robust predictions.

In addition, the emerging field of multimessenger astrophysics offers powerful complementary constraints. Observations of neutron star mergers through gravitational waves, combined with electromagnetic counterparts and precise pulsar measurements, provide independent information on the equation of state of dense matter. The synergy between heavy-ion physics and astrophysics is therefore becoming increasingly important, as both fields contribute to a unified understanding of QCD matter under extreme conditions.

Ultimately, the most important outcome of this research program may not be the identification of a single critical point in the QCD phase diagram, although such a discovery would be of great significance. Rather, the long-term goal is to construct a quantitatively constrained and internally consistent map of QCD matter across a wide range of temperature, baryon density, and isospin asymmetry. Such a map would integrate insights from experiment, theory, and observation, providing a comprehensive picture of how strongly interacting matter behaves in all its forms. In this sense, the study of nuclear matter under extreme conditions represents not only a frontier of nuclear physics, but also a key step toward a deeper understanding of the fundamental properties of matter in the universe.

\subsection{Multiscale nuclear structure: shell evolution, clustering, collectivity, and the superheavy frontier}

\begin{figure*}[htbp]
\includegraphics[width=1.0\linewidth]{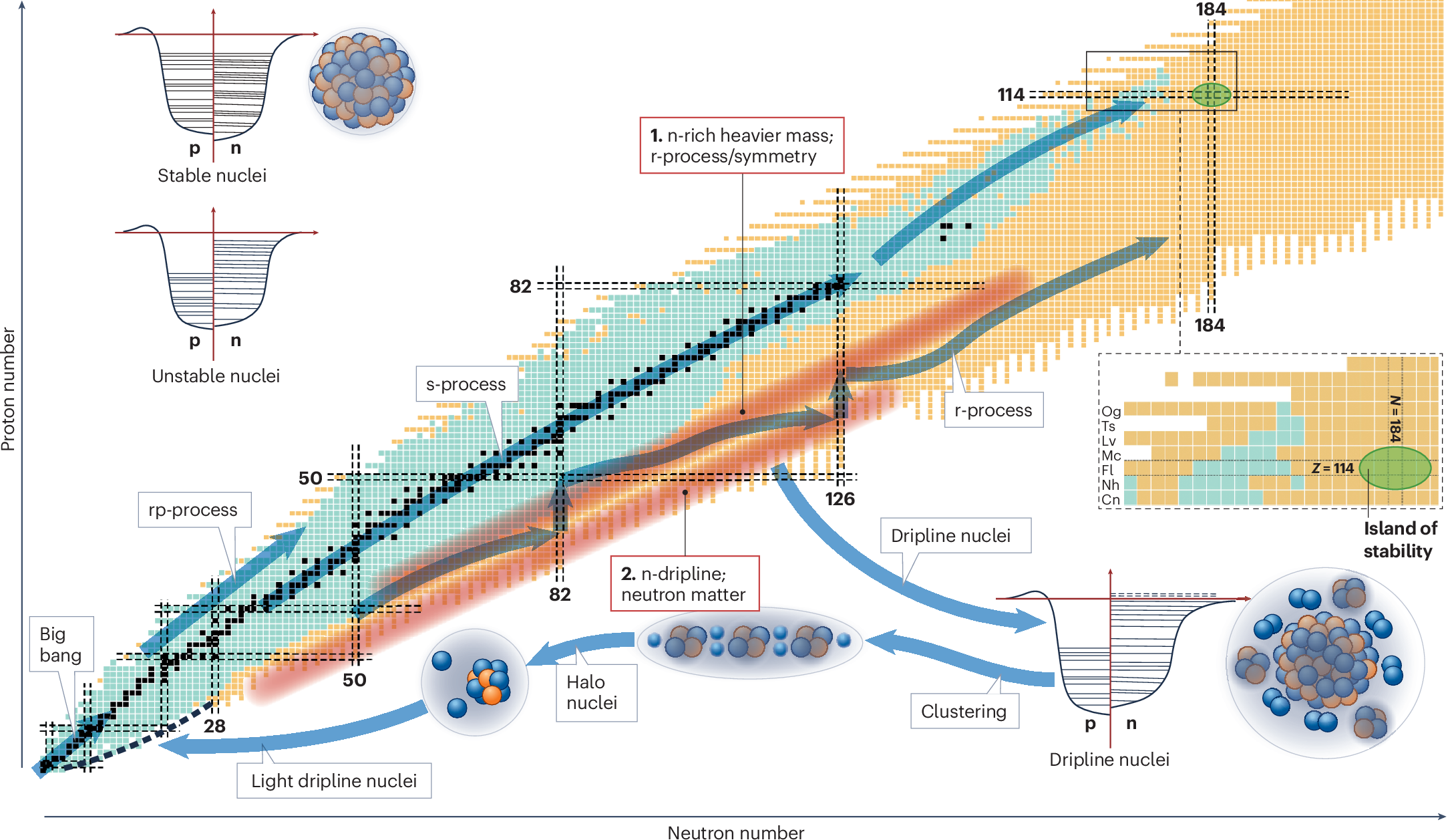}
\caption{\label{Nuclear_lanscape}
Schematic nuclear landscape for multiscale structure~\cite{Ye2025}. The figure shows the valley of stability, proton and neutron driplines, selected rare-isotope regions, halo and threshold domains, and the superheavy region.
}
\end{figure*}

Atomic nuclei are finite quantum many-body systems that display a remarkable coexistence of different structural modes across multiple length and energy scales, as shown in Fig.~\ref{Nuclear_lanscape}. From the underlying interactions among nucleons governed by the strong force, a rich hierarchy of emergent phenomena arises, including independent-particle motion in mean-field potentials, pairing correlations, collective deformation, rotational and vibrational excitations, and cluster structures. These diverse modes~\cite{Brown2001PPNP,Drut2010DFT,Moller2016,Freer2007RMP,Ye2023} are not independent; rather, they coexist and compete within the same system, often leading to intricate patterns of shell evolution, deformation, clustering, and shape coexistence~\cite{Otsuka2005PRL,Holt2013JPhysG,Zhou2026}. A central and persistent challenge for nuclear theory is therefore not merely to describe each structural feature in isolation, but to construct unified frameworks that remain predictive and internally consistent across wide regions of the nuclear chart, including both stable nuclei and those far from stability \cite{Brown2001PPNP,Navratil2016,Ekstrom2023AbInitio,Xu2024}.

The traditional shell model, which organizes nucleons into quantized energy levels within an effective mean-field potential, provides a powerful starting point for understanding nuclear structure~\cite{Brown2001PPNP}. However, it has become increasingly clear that shell structure itself is not immutable. As one moves toward exotic nuclei with large proton–neutron asymmetry, the ordering and spacing of single-particle levels can change significantly, leading to the disappearance of conventional magic numbers and the emergence of new ones. This phenomenon, known as shell evolution, is now understood to be driven in part by tensor components of the nucleon–nucleon interaction and by three-nucleon forces, which modify the effective single-particle energies in a density- and isospin-dependent manner~\cite{Otsuka2005PRL,Holt2013JPhysG}. Precision measurements of nuclear masses, excitation spectra, and electromagnetic transition strengths have provided compelling evidence for such evolution, particularly along isotopic chains near the driplines and in regions close to the $N \approx Z$ line~\cite{PhysRevLett.130.192501,Lalanne2023,Li2023,Phillips2025}.

At the same time, nuclei often exhibit strong correlations that go beyond the independent-particle picture. Pairing correlations, analogous to those in superconductors, play a crucial role in determining nuclear binding energies and excitation spectra, especially in open-shell nuclei~\cite{Dean2003Pairing}. Collective phenomena, such as quadrupole deformation and rotational motion, emerge when many nucleons coherently contribute to a common shape or dynamical mode~\cite{Moller2016}. These collective degrees of freedom are often described using models based on symmetry principles and geometric concepts, which complement more microscopic approaches. Vibrational excitations, shape coexistence, and transitional behavior between spherical and deformed configurations further enrich the phenomenology of nuclear structure~\cite{Zhou2026}.

\begin{facilitysidebar}{Sidebar 2. Rare-isotope beams, storage rings, and superheavy-element facilities}
\facilityimage{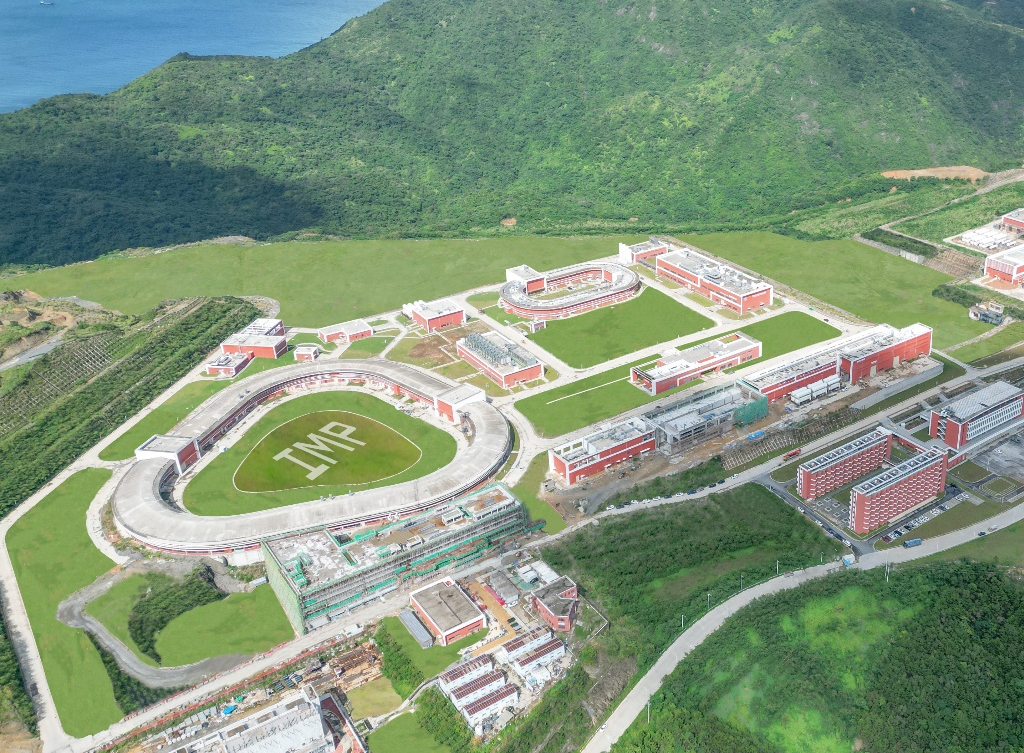}{HIAF represents the domestic heavy-ion accelerator platform that links rare isotopes, storage-ring precision, dense matter, and applications.}
\facilityentry{Weak binding and open quantum systems}
{HIAF, HIRFL-CSR, and radioactive-ion-beam programs}
{Rare-isotope beams and cooler-storage rings provide the masses, lifetimes, decay modes, electromagnetic moments, charge radii, and reaction observables needed to test shell evolution, halo formation, continuum coupling, and dripline extrapolations. HIAF strengthens this program through high-intensity heavy-ion beams, storage-ring precision, flexible reaction stations, and access to exotic nuclei relevant to both structure and astrophysics \cite{Zhou2022,HIAF,HIAF2,Brho}. The facility-level requirement is not only higher beam intensity, but also end-to-end capability: production, separation, identification, storage, detection, and rapid theory comparison.}

\facilityentry{The superheavy frontier}
{Heavy-ion fusion and separator-based experiments}
{Superheavy-element research depends on high-current heavy-ion beams, chemically and kinematically selective separators, low-background focal-plane detectors, alpha-decay and spontaneous-fission correlations, and theory-guided campaign design. Domestic HIAF-related programs therefore link the search for new isotopes to fission barriers, shell stabilization, reaction dynamics, and the long-term map of the nuclear landscape \cite{Hofmann2000RMP,PhysRevLett.134.022501,PhysRevLett.134.232501}. Planning should treat superheavy synthesis, precision mass measurements, and fission studies as one connected program rather than isolated experiments.}
\end{facilitysidebar}

In addition to these well-established modes, cluster correlations represent another important aspect of nuclear structure, particularly in light nuclei. In such systems, nucleons can organize into substructures resembling alpha particles or other light clusters, leading to configurations that are qualitatively different from mean-field descriptions. These cluster states often appear near particle emission thresholds and are closely related to resonant structures in the continuum. Understanding the interplay between clusterization and shell-model configurations remains an active area of research, with implications for both nuclear structure and nuclear reactions, including those relevant to astrophysical processes \cite{Elhatisari2015}.

Recent reviews of exotic nuclei, clustering phenomena, and modern many-body methods indicate that the field is steadily progressing toward a more unified and comprehensive description of nuclear structure across scales \cite{Ye2025,Ye2023,YeNT,Xu2024,Stroberg2021,Miyagi2022}. On the microscopic side, significant advances have been achieved through \textit{ab initio} approaches, which aim to solve the nuclear many-body problem starting from realistic nucleon–nucleon and three-nucleon interactions derived from chiral effective field theory. Methods such as the no-core shell model, coupled-cluster theory, and in-medium similarity renormalization group have demonstrated impressive success in describing light and medium-mass nuclei with controlled approximations. At the same time, many-body perturbation theory and valence-space effective interactions provide practical tools for extending these insights to heavier systems.

A key development in recent years is the increasing importance of continuum effects in nuclear structure theory. For nuclei near the driplines, where binding energies are small and particle emission thresholds are low, the coupling between bound states and the continuum becomes essential. This leads to phenomena such as halo structures, resonances with finite lifetimes, and enhanced reaction coupling. Continuum-aware approaches, including the use of Berggren bases and complex-energy formalisms, allow for a consistent treatment of bound, resonant, and scattering states within a unified framework \cite{Xu2024,Navratil2016,Berggren1968,Michel2021,Okolowicz2003PhysRept}. These methods are crucial for describing weakly bound systems and for connecting structure and reaction dynamics in a consistent manner.

The multiscale nature of nuclear structure becomes particularly striking in the region of superheavy nuclei. In this regime, the existence of nuclei is governed by a delicate balance between several competing effects. The strong Coulomb repulsion between the large number of protons tends to destabilize the nucleus and drive it toward fission. At the same time, shell effects arising from quantized single-particle motion can provide additional binding and stabilize certain configurations. Deformation and shape coexistence further complicate the picture, as nuclei may adopt non-spherical shapes that lower their energy. The concept of an “island of stability,” corresponding to superheavy nuclei with relatively long lifetimes due to favorable shell closures, has long been a central goal of experimental and theoretical efforts.

However, the modern perspective on superheavy nuclei goes beyond the search for a few magic numbers  \cite{Hofmann2000RMP}. Instead, the focus has shifted toward understanding the full landscape of structural and dynamical properties in this extreme regime. This includes the evolution of shell corrections with increasing proton number, the role of deformation and triaxiality, the competition between different fission pathways, and the impact of dissipative dynamics on fission barriers and lifetimes. Recent experimental advances have enabled the synthesis and study of new superheavy isotopes, albeit with very short lifetimes, providing valuable data on decay modes, excitation energies, and fission properties \cite{PhysRevLett.134.022501,PhysRevLett.134.232501}. These measurements offer critical tests of theoretical models and help constrain predictions in regions that remain experimentally inaccessible.

On the theoretical side, global nuclear mass models and advanced fission theories remain indispensable tools for exploring the superheavy region and guiding experimental searches \cite{Moller2016,Bender2020}. Approaches based on energy density functionals, macroscopic–microscopic models, and self-consistent mean-field theories provide complementary perspectives on nuclear binding and deformation. At the same time, quantifying theoretical uncertainties has become an increasingly important objective, particularly when extrapolating to unknown regions of the nuclear chart. Efforts to improve predictive power include the incorporation of more accurate interactions, better treatment of correlations, and systematic benchmarking against experimental data.

In summary, the study of multiscale nuclear structure represents a central pillar of modern nuclear physics, encompassing a wide range of phenomena from single-particle motion to collective dynamics and clustering \cite{Brown2001PPNP,Freer2007RMP,Ye2023}. The challenge of achieving a unified description across these scales is being met through a combination of experimental advances and theoretical innovations. Nowhere is this challenge more evident than in the exploration of superheavy nuclei, where the limits of nuclear existence are tested and the interplay of fundamental forces reaches its most extreme form. Continued progress in this area will not only deepen our understanding of nuclear structure, but also shed light on the fundamental principles governing complex quantum many-body systems.

\subsection{Exotic nuclei, open quantum systems, and new reaction mechanisms}

\begin{figure*}[htb]
\floatbox[{\capbeside\thisfloatsetup{capbesideposition={right,top},capbesidewidth=0.2\textwidth}}]{figure}[\FBwidth]
{\caption{Schematic nuclear landscape for open quantum systems. The figure emphasizes how shell evolution, clustering, deformation, continuum coupling, and rare decay modes become increasingly intertwined toward the limits of nuclear existence. {Original schematic prepared by the author.}}\label{fig:nuclear-landscape-open-systems}}
{\includegraphics[width=0.75\textwidth]{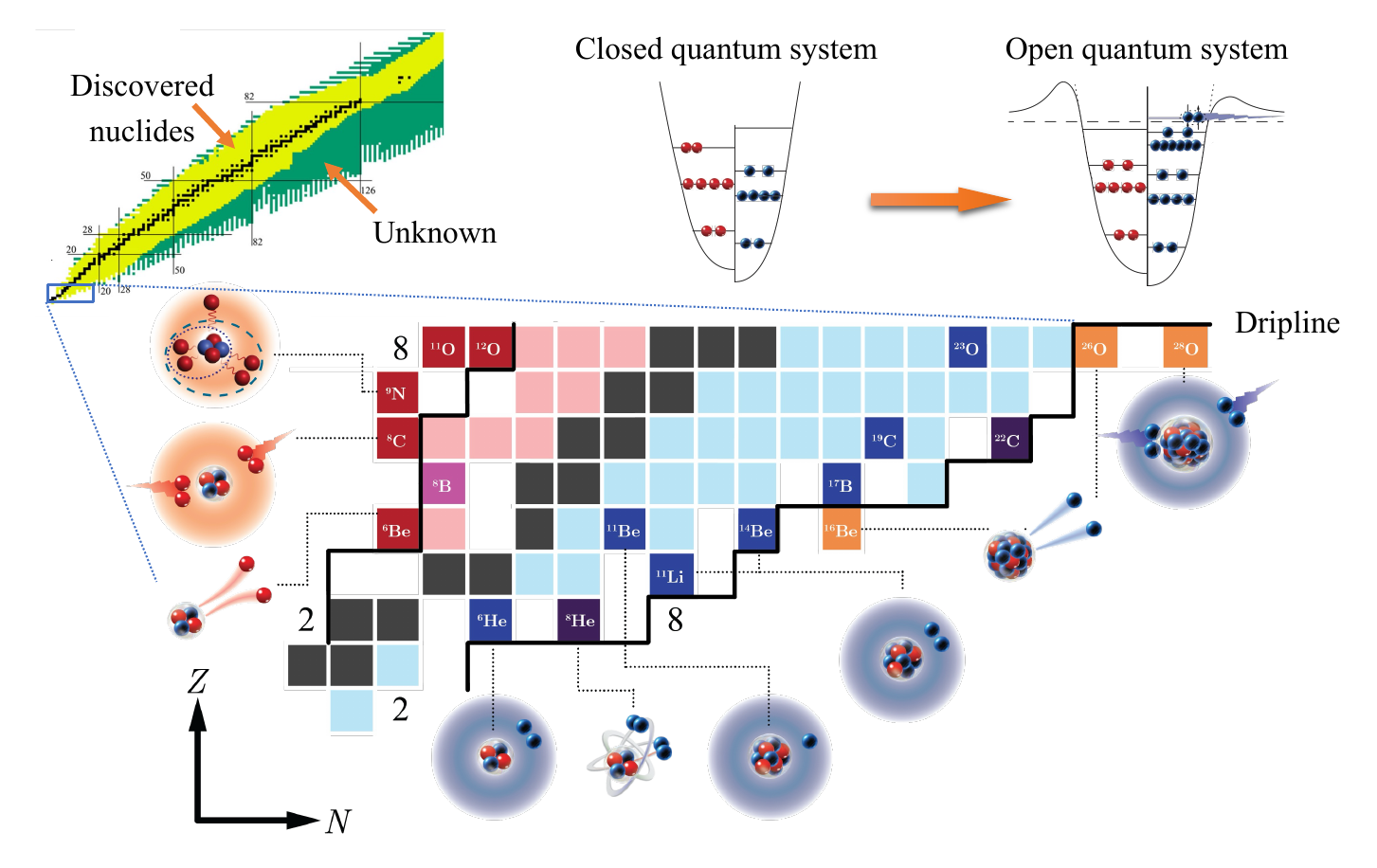}}
\end{figure*}

Nuclei located near the proton and neutron driplines represent some of the most extreme and informative systems in nuclear physics. Unlike stable nuclei, which can often be approximated as closed quantum systems with well-defined bound states, these weakly bound or unbound nuclei are more appropriately described as open quantum systems (see Fig.~\ref{fig:nuclear-landscape-open-systems}). Their defining characteristics—low separation energies, proximity to particle emission thresholds, and strong coupling to the scattering continuum—lead to qualitative modifications of their structural and dynamical properties. As a result, conventional theoretical frameworks based solely on bound-state approximations become inadequate, and a unified treatment that incorporates both structure and reaction dynamics is required.

The open quantum nature of exotic nuclei manifests itself in several fundamental ways \cite{Okolowicz2003PhysRept}. First, the coupling between discrete states and the continuum can significantly alter shell structure, leading to shifts in single-particle energies, modified level ordering, and even the disappearance or emergence of shell closures \cite{Otsuka2005PRL,Brown2001PPNP,Fossez2017}. Second, the spatial extension of weakly bound nucleons can become exceptionally large, giving rise to halo structures in which one or two nucleons occupy diffuse orbitals that extend far beyond the nuclear core. Third, decay patterns become more complex, often involving direct emission into the continuum, multi-particle correlations, and competition between different decay channels. Finally, reaction observables, such as cross sections and momentum distributions, become highly sensitive to continuum coupling and asymptotic wave-function behavior.

Because of these features, the traditional boundary between nuclear structure and reaction theory becomes increasingly blurred in exotic systems. Observables that were once interpreted primarily in terms of internal structure must now be understood in the context of coupling to external channels, while reaction processes themselves can reveal detailed information about underlying correlations and configurations. This interplay has made the physics of exotic nuclei one of the most compelling laboratories for studying emergent behavior in finite quantum many-body systems \cite{Ye2025,YeNT}. In particular, it provides a unique opportunity to test how fundamental interactions give rise to new forms of organization when the stabilizing influence of strong binding is reduced.

A number of distinctive phenomena have been identified in this context. Halo nuclei, such as those found near the neutron dripline, exhibit extended matter distributions and narrow momentum profiles, reflecting the weak binding and low angular momentum of valence nucleons \cite{Jonson2004,Jensen2004,Hansen1987AnnRev,Tanihata2013PPNP}. Threshold phenomena, including the appearance of resonances near particle emission energies, highlight the importance of continuum coupling and the role of near-threshold dynamics. Soft excitation modes, such as pygmy resonances, arise from oscillations of weakly bound nucleons against a more tightly bound core. Mirror asymmetries between proton-rich and neutron-rich systems reveal the interplay between Coulomb effects and nuclear forces, while continuum-induced correlations can generate unexpected structures and decay patterns. Unusual decay modes, including one- and two-nucleon radioactivity, further illustrate how the loss of stability leads to qualitatively new behavior \cite{Michel2006,Aksyutina2008,Kondo2016,Webb2018,Thomas1952,Hoff2020,Algora2025,Michel2010,Zhang2022MirrorOxygen,Campbell2024,Charity2023,Xu2025a}.

Recent topical reviews have emphasized additional layers of complexity, particularly in systems that combine weak binding with deformation or isospin asymmetry \cite{ZhouNT,LinNT}. Deformed halo nuclei, for example, challenge the traditional separation between collective and single-particle motion, as the extended halo structure must be described within a deformed mean field that itself evolves with the underlying configuration. Similarly, low-energy reaction dynamics in proton-rich systems are strongly influenced by Coulomb barriers and resonance structures, requiring careful treatment of both nuclear and electromagnetic interactions. These developments underscore the necessity of frameworks that treat structure and reactions on equal footing, rather than as separate domains.

Another important aspect of exotic nuclei research is the exploration of the limits of nuclear existence. Determining the location of the proton and neutron driplines, as well as the total number of bound nuclei, remains an active area of investigation \cite{Erler2012NuclearLandscape,Forssen2013}. Both experimental advances in radioactive beam facilities and theoretical developments in global nuclear models contribute to this effort. Precision mass measurements, particularly near the proton dripline, have provided new insights into binding energies, decay thresholds, and shell evolution, often challenging existing theoretical predictions \cite{PhysRevLett.133.222501,PhysRevLett.135.012501,Campbell2024,ZhouNP}. These measurements are essential for refining models of nuclear structure and for understanding how stability is lost as one approaches the edges of the nuclear landscape.

From a theoretical perspective, the description of exotic nuclei requires methods that explicitly incorporate the open quantum nature of the system. Among the most powerful approaches are those based on the Berggren ensemble, which extends the conventional basis of bound states to include resonant and scattering states in the complex energy plane. This formalism underlies the Gamow shell model and related continuum shell-model approaches, which provide a unified framework for treating bound, resonant, and continuum configurations on the same footing \cite{Berggren1968,Michel2021,Okolowicz2003PhysRept,Aoyama2006,Hu2020}. These methods allow for a consistent description of decay widths, resonance properties, and continuum coupling effects, making them particularly well suited for systems near the driplines.

In parallel, cluster models and coupled-channel approaches remain indispensable for describing reaction observables and asymptotic correlations \cite{Ikeda1968PTPS,Freer2007RMP,Yahiro2012PPNP}. These frameworks are especially important for processes involving few-body dynamics, such as nucleon emission and breakup reactions \cite{Alt1967,Barker2003,Alvarez2008}, where the relative motion of clusters or fragments plays a dominant role. By combining microscopic structure input with reaction theory, these approaches enable detailed comparisons with experimental data and provide insight into the mechanisms underlying observed phenomena \cite{Yahiro2012PPNP,Navratil2016}.

Two-proton radioactivity serves as a paradigmatic example of how open quantum system concepts have transformed our understanding of nuclear decay. Initially proposed by Ref.~\cite{Goldansky1960} as a possible decay mode for proton-rich nuclei beyond the dripline, it has since been experimentally observed and extensively studied. Theoretical and experimental work has revealed a rich interplay between decay energy, proton–proton correlations, and emission mechanisms. In some cases, the decay proceeds as a prompt, correlated two-proton emission, while in others it follows a sequential pathway through intermediate states. The analysis of such processes requires a full three-body treatment, including the asymptotic behavior of the wave function and the influence of continuum coupling \cite{Pfutzner2023,Zhou2022a,Blank2008,Blank2005,Ascher2011,Brown2014,Miernik2007b,Blank2010}. These studies not only deepen our understanding of nuclear decay, but also provide stringent tests of theoretical models in the open quantum regime.

More broadly, the study of exotic nuclei raises a fundamental question: how can the diverse and often unexpected phenomena observed near the limits of stability be organized into a coherent and predictive theoretical framework? While it is now clear that new forms of structure and dynamics emerge in these systems, the challenge lies in identifying the underlying principles that govern their behavior and in developing models that can reliably extrapolate to unknown regions. Achieving this goal will require continued integration of structure and reaction theory, improved treatment of continuum effects, and close collaboration between experiment and theory.

In this sense, exotic nuclei are not merely a peripheral extension of nuclear physics, but a central testing ground for its most fundamental concepts. By pushing the boundaries of stability and exploring the role of openness, they provide critical insights into the nature of quantum many-body systems and the limits of nuclear existence. {In summary, the defining task is to treat structure, resonances, decay, and reactions within one uncertainty-quantified open-quantum-system framework, with rare-isotope measurements and continuum theory developed in direct feedback.}

\subsection{Nuclear astrophysics and the origin of the elements}

Nuclear astrophysics seeks to establish a fundamental connection between the microscopic properties of atomic nuclei and the macroscopic evolution of the universe, particularly in relation to the origin and abundance of the chemical elements \cite{B2FH1957}. At its core, the field addresses how nuclear reactions, decay processes, and nuclear structure properties govern the synthesis of elements in stars and stellar explosions. While the broad conceptual framework of nucleosynthesis has been established over decades of research, many of the underlying details remain uncertain, especially for processes involving unstable nuclei far from stability. The central scientific challenge is therefore to quantitatively link specific astrophysical environments with the relevant nuclear inputs, including nuclear masses, weak interaction rates, reaction cross sections, and decay lifetimes, which collectively determine the flow of nucleosynthesis pathways~\cite{Kolos2022NuclearDataNeeds,Cowan2021RProcess}.

\begin{figure*}[!htb]
    \includegraphics[width =1\linewidth]{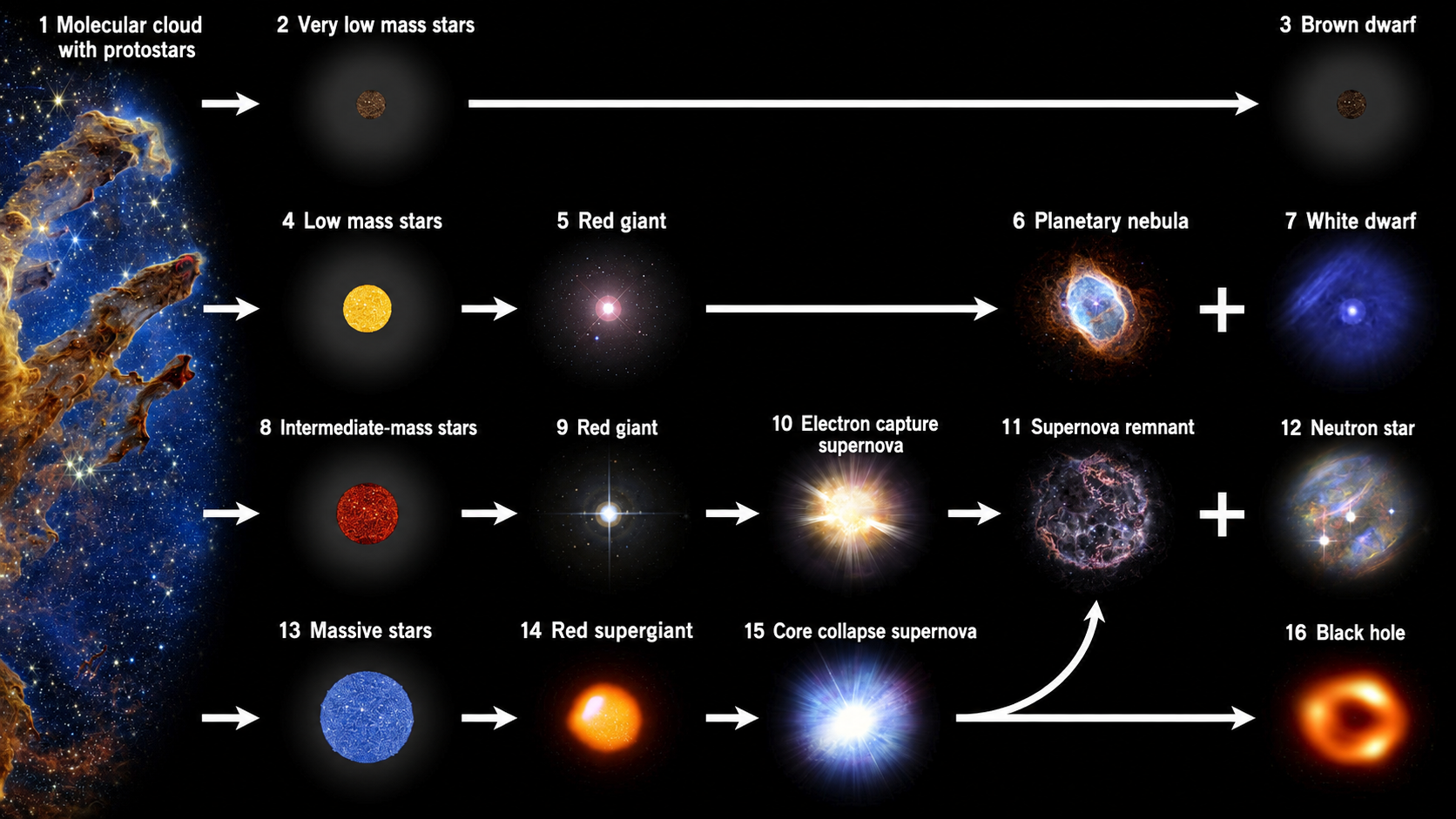}
    \caption{
Stellar evolution for low- and intermediate-mass and massive stars. Recent research suggests that intermediate-mass stars can also evolve
toward thermal explosion and a white dwarf. Adapted from Ref.~\cite{nupecc2024lrp}.
}
\label{fig:stellar_evolution}
\end{figure*}

In stellar environments, nucleosynthesis proceeds through a variety of mechanisms depending on temperature, density, and composition (see Fig.~\ref{fig:stellar_evolution}). Hydrostatic burning phases in stars, such as hydrogen burning through the proton-proton chains and CNO cycles, as well as helium, carbon, and silicon burning, are responsible for producing many of the light and intermediate-mass elements~\cite{B2FH1957,ChenYJNST,GuoPPNP,CJPL}. These processes occur under relatively stable conditions and can often be described using well-established reaction networks. By contrast, the synthesis of many elements beyond iron requires more extreme environments. Explosive scenarios, including core-collapse supernovae and neutron-star mergers, provide high temperatures, intense neutron fluxes, and rapid expansion conditions for heavy-element nucleosynthesis, most notably the rapid neutron-capture process ($r$-process), while the slow neutron-capture process ($s$ process) and proton-rich nucleosynthesis channels complete the broader picture of heavy-element production~\cite{B2FH1957,Cowan2021RProcess}.

A defining feature of many of these nucleosynthesis pathways is their reliance on nuclei that are highly unstable and often experimentally inaccessible. For example, the $r$-process involves rapid neutron captures on neutron-rich isotopes far from the valley of stability, followed by beta decay back toward stable nuclei. The exact path of the $r$-process, as well as the resulting abundance patterns, depends sensitively on nuclear masses, neutron separation energies, beta-decay rates, and fission properties of these exotic nuclei. Similarly, in proton-rich environments, processes such as the rapid proton capture process ($rp$-process) and the $\nu p$-process involve sequences of proton captures and beta decays on unstable isotopes near the proton dripline. In all cases, the lack of precise nuclear data introduces significant uncertainties into astrophysical models, making it difficult to draw definitive conclusions about the origin of specific elements.

Recent reviews have highlighted substantial progress in both experimental capabilities and astrophysical modeling \cite{LiuNST2024,LiuNT}. On the experimental side, advances in rare-isotope beam facilities, storage rings, and detection technologies have enabled increasingly precise measurements of key nuclear properties. Techniques such as recoil separators, transfer reactions, and indirect methods (e.g., Coulomb dissociation and Trojan horse approaches) have been developed to access reaction rates that are otherwise difficult or impossible to measure directly. Underground laboratories, which provide low-background environments, are particularly important for studying reactions with very low cross sections, such as those relevant to stellar hydrogen burning. In China, the China Jinping Underground Laboratory (CJPL) hosts a new generation of deep-underground nuclear astrophysics experiments, taking advantage of its world-leading rock overburden to achieve an ultra-low cosmic-ray background environment \cite{CJPL}. This unique setting enables direct measurements of key nuclear reaction cross sections at stellar energies, which are otherwise inaccessible on the surface. By combining high-intensity accelerators with ultra-sensitive detection systems, CJPL aims to reduce uncertainties in reaction rates that govern stellar evolution and nucleosynthesis. These efforts play a crucial role in advancing our understanding of the origin of elements and the energy production mechanisms in stars.

\begin{facilitysidebar}{Sidebar 3. Underground and stellar-reaction facilities}
\facilityimage{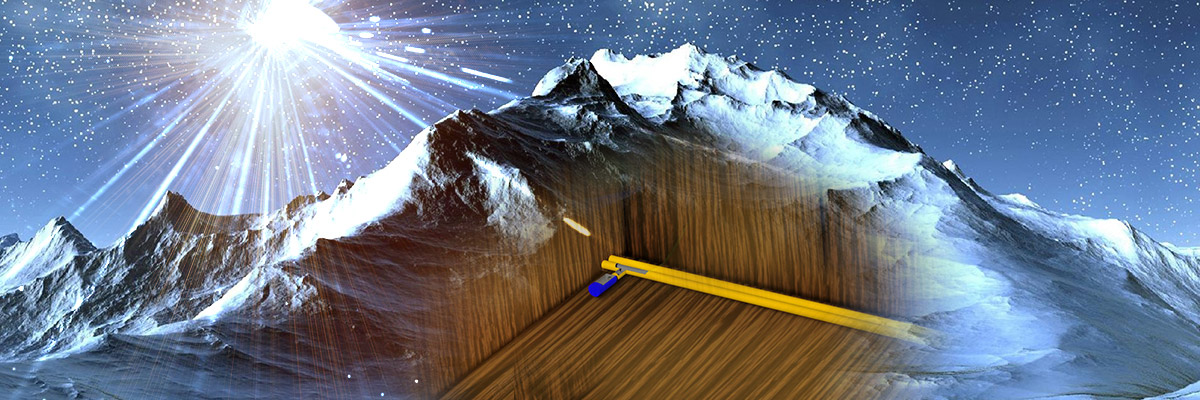}{A schematic plot for the China Jinping underground laboratory (CJPL).}
\facilityentry{Low-background stellar burning}
{CJPL and JUNA}
{Deep-underground accelerators suppress cosmic-ray backgrounds and enable direct measurements of extremely small cross sections at stellar energies. CJPL/JUNA therefore anchors the domestic program for hydrogen burning, helium burning, breakout reactions from the CNO cycles, and reaction-rate normalization in stellar-evolution models \cite{CJPL,ChenYJNST,LiuNST2024,LiuNT}. The critical technical ingredients are high-current stable beams, ultra-low-background gamma and charged-particle detectors, radiopure materials, and target systems whose degradation and contamination can be controlled at the percent level.}

\facilityentry{Unstable-nucleus astrophysics}
{Rare-isotope beams, recoil separators, and storage rings}
{Explosive nucleosynthesis requires masses, beta-decay properties, neutron-emission probabilities, fission yields, and reaction rates for short-lived nuclei. Storage-ring and rare-isotope techniques complement underground direct measurements by extending the experimental reach toward the $r$-, $rp$-, and $\nu p$-process paths \cite{Glorius2023StorageRingAstro,PhysRevLett.134.082701}. A mature program therefore requires coordinated access to underground laboratories, rare-isotope beams, recoil separators, storage rings, and astrophysical simulations so that each measured rate can be propagated into abundance uncertainties.}
\end{facilitysidebar}

Representative examples of recent progress include new evidence for breakout pathways from the CNO cycles in the first generation of stars, which can significantly alter the production of heavier elements in early stellar environments \cite{Zhang2022}. In addition, improved storage-ring techniques have opened new possibilities for studying reactions involving short-lived isotopes, allowing for direct measurements in previously inaccessible regions of the nuclear chart \cite{PhysRevLett.134.082701}. These experimental advances are essential for reducing uncertainties in reaction rates and for providing reliable input to astrophysical models.

On the theoretical side, nuclear astrophysics has benefited greatly from the rapid development of multimessenger astronomy. Observations of gravitational waves from neutron star mergers, combined with electromagnetic signals across the spectrum, have provided compelling evidence for the production of heavy elements via the $r$-process in such events. These observations have significantly strengthened the connection between nuclear physics and astrophysical phenomena, offering new constraints on the properties of dense matter and the conditions under which nucleosynthesis occurs. At the same time, theoretical models of neutrino interactions in supernovae and mergers have highlighted the importance of neutrino-induced processes in shaping nucleosynthesis pathways, particularly in determining the neutron-to-proton ratio in ejecta \cite{Cowan2021RProcess}.

Despite these advances, several key questions remain open. The relative contributions of different astrophysical sites to the observed abundances of heavy elements are still under debate, as are the detailed mechanisms of element production in specific environments. The origin of certain rare isotopes, particularly the so-called $p$-nuclei, continues to challenge existing models, as their production requires specific conditions that are not yet fully understood. Furthermore, the interplay between nuclear physics uncertainties and astrophysical modeling complicates efforts to extract precise information from observational data.

The frontier challenge in nuclear astrophysics has thus evolved beyond the identification of individual nucleosynthesis sites or processes. Instead, the goal is to construct a quantitatively consistent and predictive framework that integrates multiple environments, processes, and observational constraints. Achieving this requires a comprehensive approach that combines improved nuclear data, advanced reaction modeling, and robust statistical methods for uncertainty quantification and propagation. In particular, sensitivity studies and Bayesian inference techniques are increasingly being used to identify the most critical nuclear inputs and to constrain models using observational data.

A key component of this effort is the continued development of experimental programs targeting unstable nuclei. As next-generation rare-isotope facilities come online and existing storage rings are upgraded, it will become possible to measure nuclear properties with unprecedented precision and to explore previously inaccessible regions of the nuclear chart. These measurements must be closely coordinated with theoretical efforts to ensure that the resulting data can be effectively incorporated into astrophysical models.

Equally important is the integration of nuclear physics with astronomical observations \cite{Cowan2021RProcess}. High-resolution spectroscopy of stars, observations of supernova light curves, and multimessenger signals from compact-object mergers all provide valuable information about nucleosynthesis processes. By combining these observations with detailed nuclear models, it becomes possible to test theoretical predictions and to refine our understanding of the origin of the elements.

In this context, nuclear astrophysics is increasingly moving toward coordinated, interdisciplinary research programs in which experiment, theory, and observation are developed in tandem. Rather than treating these components as separate endeavors, the field is recognizing the importance of designing experiments and simulations that directly address astrophysical questions, while also using observational data to guide and constrain nuclear physics research.

In summary, nuclear astrophysics represents a dynamic and rapidly evolving field that lies at the intersection of nuclear physics, astrophysics, and cosmology. By linking microscopic nuclear properties to the large-scale structure and evolution of the universe, it provides a powerful framework for understanding the origin of the elements. Continued progress will depend on the integration of experimental advances, theoretical developments, and observational insights, ultimately leading to a more complete and quantitatively accurate picture of how the elements that make up our world were formed.

\section{Enabling methodologies: theory, facilities, precision instrumentation, and data-driven methods}

\subsection{Modern nuclear theory as a bridge across subfields}

\begin{figure*}[htb]
\floatbox[{\capbeside\thisfloatsetup{capbesideposition={right,top},capbesidewidth=0.2\textwidth}}]{figure}[\FBwidth]
{\caption{
    Schematic illustration of modern nuclear theory as an interoperable framework connecting traditionally separated subfields. 
    Ab initio many-body methods, effective interactions, continuum and open-quantum-system approaches, reaction theory, and uncertainty quantification provide complementary theoretical components for describing nuclear structure, reactions, astrophysical processes, dense matter, and fundamental symmetries within a unified predictive framework. {Original schematic prepared by the author.}
    }\label{fig:Nuclear_theory}}
{\includegraphics[width=0.75\textwidth]{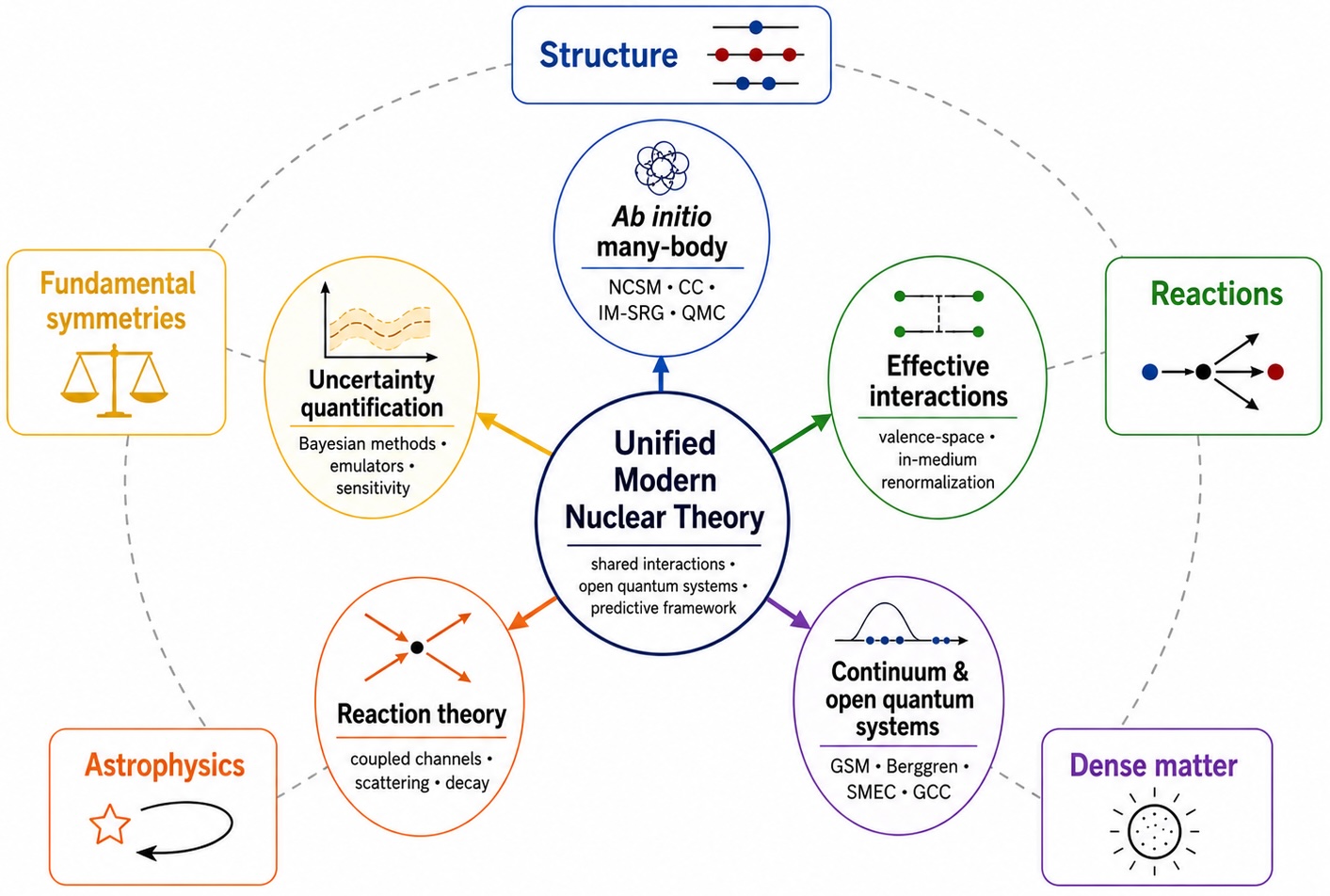}}
\end{figure*}

A defining characteristic of contemporary nuclear science is the increasing convergence of theoretical frameworks that were historically developed for distinct subfields. Techniques originally designed for light nuclei, nuclear structure, nuclear reactions, and nuclear astrophysics are now being integrated into a more unified theoretical ecosystem. This convergence is not merely a matter of methodological convenience; it reflects a deeper recognition that the same underlying nuclear interactions govern phenomena across vastly different energy scales, density regimes, and nuclear environments. As a result, modern nuclear theory is progressively evolving into a shared infrastructure that supports multiple frontiers of research, from the spectroscopy of exotic nuclei to the equation of state of dense matter and the interpretation of astrophysical observations (see Fig.~\ref{fig:Nuclear_theory}).

Among the most important developments in this direction are \textit{ab initio} many-body methods, which aim to describe nuclear systems starting from realistic nucleon–nucleon and three-nucleon interactions derived from quantum chromodynamics via effective field theory \cite{Drut2010DFT,Ekstrom2023AbInitio}. These approaches include methods such as the no-core shell model, coupled-cluster theory, in-medium similarity renormalization group (IM-SRG), and quantum Monte Carlo techniques~\cite{Navratil2016,Xu2024,Adams2021NQS,Fore2023NQSNeutronMatter,Zhang2026HyperNQS,Zhang2026ExcitedNQS}. A key strength of these frameworks is that they provide a systematically improvable expansion with controlled approximations, allowing for quantifiable uncertainties in calculated observables. This feature is particularly important in an era where nuclear theory is increasingly expected to provide not only qualitative explanations but also high-precision predictions.

In parallel, significant progress has been made in the development of effective interactions and renormalization techniques tailored to specific nuclear environments. In-medium renormalization methods, for example, allow complex many-body correlations to be incorporated into reduced valence-space descriptions, enabling accurate calculations of medium-mass and heavy nuclei that would otherwise be computationally intractable. When combined with modern computational resources, these approaches have substantially extended the reach of microscopic nuclear theory beyond the lightest systems, bridging the long-standing gap between fundamental interactions and nuclear phenomenology~\cite{Navratil2016,Stroberg2021,Miyagi2022,Xu2024,Fore2023NQSNeutronMatter,Zhang2026HyperNQS,Zhang2026ExcitedNQS}.

Closely related to these computational advances, quantum information science is opening a further frontier for nuclear theory. Its relevance goes beyond the possible use of quantum processors as faster numerical devices: by representing many-body evolution through entanglement, unitary operations, and measurement, quantum information reformulates aspects of nuclear dynamics in a language that is natural to both quantum hardware and the underlying physics. This viewpoint is particularly compelling for problems in which real-time response functions, scattering amplitudes, or strongly entangled wave functions remain difficult to access with conventional classical algorithms. Early demonstrations, including quantum computations of light nuclei, neutrino--nucleus response functions, and resource estimates for nuclear effective field theories, have shown how nuclear Hamiltonians can be mapped onto quantum circuits and explored within hybrid quantum--classical workflows~\cite{Dumitrescu2018CloudNucleus,Roggero2020NeutrinoNucleusQC,Bauer2023QuantumSimulationHEP,Savage2024QuantumComputingNP,Watson2023NuclearEFTQC}. In parallel, quantum-information measures are beginning to provide new diagnostics of nuclear correlations, from cluster degrees of freedom and femtometer-scale interference to spin entanglement in nucleon--nucleon scattering~\cite{Ma2023FermiDoubleSlit,Xu2026ClusterEntanglement,Shen2025Entanglement}.

A particularly important conceptual advance is the recognition that continuum coupling is not a peripheral correction, but a fundamental aspect of nuclear dynamics in many regimes. In weakly bound or unbound systems, the proximity of particle emission thresholds leads to strong coupling between discrete bound states and the scattering continuum. This coupling affects not only decay properties and resonance widths, but also the very structure of nuclear states, including level ordering, spectroscopic strength distributions, and spatial extensions of wave functions. As a result, traditional bound-state approaches must be extended to include open quantum system dynamics whenever thresholds or resonant channels play a significant role.

Frameworks such as the Gamow shell model, Berggren ensemble methods, and complex-energy continuum shell models provide a unified theoretical language for treating bound states, resonances, and scattering states on equal footing. These approaches are particularly powerful for describing exotic nuclei near the driplines, where weak binding and continuum effects dominate the physics. In addition, coupled-channel methods and reaction-theory formalisms remain essential for connecting microscopic structure information to experimentally accessible observables, such as cross sections, momentum distributions, and decay correlations \cite{Okolowicz2003PhysRept,Michel2021,Aoyama2006,Hu2020}. The integration of these techniques represents a major step toward a truly unified description of nuclear systems as open quantum many-body systems.

One of the central insights emerging from these developments is that continuum effects are not limited to a small subset of exotic nuclei, but rather constitute a structural ingredient of nuclear theory more broadly. Even in stable nuclei, excited states near particle emission thresholds, giant resonances, and reaction processes involving weakly bound channels require an explicit treatment of coupling to the continuum. This realization has led to a gradual shift in perspective: instead of treating structure and reactions as separate domains connected only at the level of observable quantities, modern nuclear theory increasingly views them as different manifestations of a single underlying open quantum system dynamics.

Another important lesson is that the most powerful theoretical frameworks for future nuclear science will not be defined solely by their accuracy in isolated calculations, but by their interoperability across different domains. In this context, interoperability refers to the ability of theoretical methods to connect nuclear structure, reaction dynamics, astrophysical processes, and uncertainty quantification within a consistent and transferable framework. For example, a theoretical description of nuclear masses should be compatible with reaction models used in astrophysical network calculations, while also being consistent with spectroscopic data and electroweak transition strengths. Achieving such consistency requires not only improved interactions and many-body methods, but also standardized interfaces between different computational approaches.

Uncertainty quantification has emerged as a key component of this interoperable framework. As nuclear theory becomes increasingly precise, it is essential to accompany predictions with reliable estimates of theoretical uncertainties arising from truncations in effective field theories, many-body approximations, and model assumptions \cite{Dobaczewski2014,Furnstahl2015,Melendez2019}. Bayesian statistical methods, emulator techniques, and sensitivity analyses are now being widely adopted to systematically propagate uncertainties from microscopic inputs to macroscopic observables \cite{Hoeting1999,Neufcourt2020b,Cook2025}. This is particularly important in applications to nuclear astrophysics and fundamental symmetries, where theoretical uncertainties can significantly impact the interpretation of experimental data.

The broader implication of these developments is that advances originally motivated by specific problems—such as the structure of dripline nuclei or the dynamics of light-ion reactions—are now reshaping much wider areas of nuclear physics. Techniques developed for treating weakly bound systems are being applied to reaction theory, astrophysical modeling, and even the study of dense matter. Conversely, insights from astrophysics and heavy-ion physics are feeding back into nuclear structure theory, motivating improved descriptions of nuclear interactions under extreme conditions.

In this evolving landscape, modern nuclear theory is increasingly functioning as a unifying framework that bridges traditionally separated subfields. Rather than operating as independent domains, nuclear structure, reactions, astrophysics, and fundamental interaction theory are becoming interconnected components of a single, integrated scientific enterprise. This transformation is being driven by both conceptual advances and computational capabilities, and it is expected to continue shaping the field in the coming decade.

Ultimately, the development of such a unified theoretical infrastructure is essential for addressing the most fundamental questions in nuclear science. These include the origin of nuclear mass and structure, the limits of nuclear existence, the nature of nuclear matter under extreme conditions, and the astrophysical origin of the elements. By providing a common language and computational foundation across subfields, modern nuclear theory is enabling a new level of coherence and predictive power in the study of strongly interacting many-body systems.

\subsection{Accelerator complexes, storage rings, light sources, and detector systems}

\begin{figure}[!htb]
    \includegraphics[width =1\linewidth]{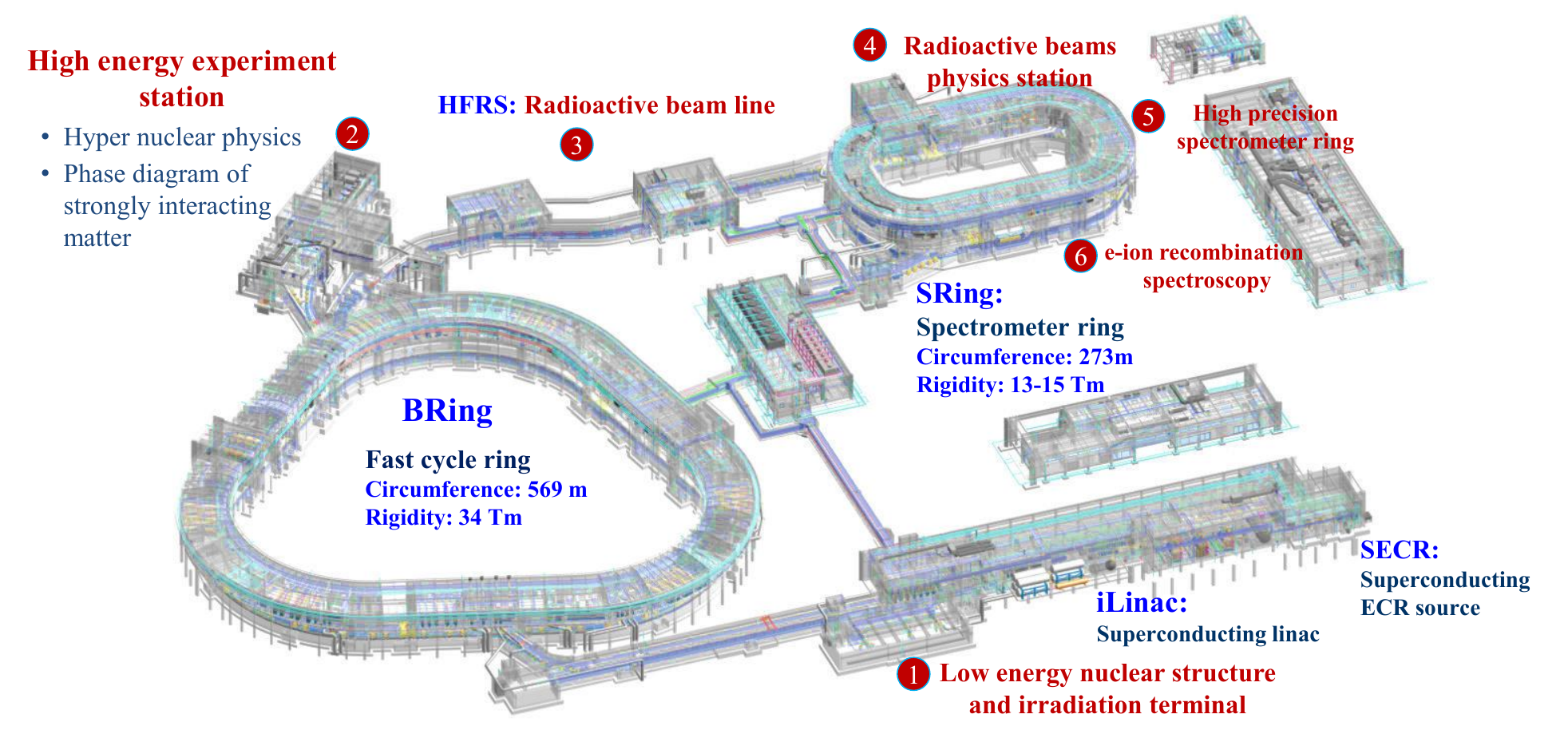}
    \caption{{HIAF schematic plot in Huizhou, Guangdong, China~\cite{HIAF}.}}
    \label{HIAF}
\end{figure}

Major advances in experimental nuclear science are inseparable from the continuous evolution of large-scale facility capabilities. In modern nuclear physics, accelerators and associated experimental infrastructures are no longer passive instruments that simply deliver particles to targets; rather, they define the very boundaries of observable phenomena. The reachable regions of the nuclear chart, the precision of extracted observables, and even the types of physical questions that can be meaningfully addressed are all fundamentally constrained and enabled by accelerator complexes, storage rings, recoil separators, neutron and photon sources, and advanced detector systems \cite{Beceiro2015,Mauss2019,Roger2018}. As nuclear science moves deeper into the domain of rare isotopes and extreme conditions, the role of integrated facility design becomes increasingly central.

Heavy-ion accelerator complexes form the backbone of much of contemporary nuclear research (Fig.~\ref{HIAF}). These facilities provide beams spanning a wide range of energies and ion species, enabling studies from near-barrier nuclear structure to relativistic heavy-ion collisions. In particular, next-generation facilities emphasize both intensity and flexibility, allowing the production of rare isotopes far from stability and the exploration of highly asymmetric nuclear systems. The development of radioactive ion beam (RIB) technology has been transformative in this regard, opening access to nuclei with extreme neutron-to-proton ratios that were previously unreachable. These isotopes are essential for understanding shell evolution, nuclear astrophysics pathways, and the limits of nuclear existence.

In this global context, recent progress in Chinese accelerator infrastructure provides a particularly clear illustration of how facility development drives scientific opportunity. The commissioning of new beamlines and the continued construction and optimization of HIAF significantly enhance national capabilities in rare-isotope production, nuclear structure studies, nuclear astrophysics, and high-energy-density physics \cite{Wang2022,Zhou2022,HIAF}. With its combination of high beam intensity, wide energy range, and advanced separation systems, HIAF is designed to enable experiments that probe previously inaccessible regions of the nuclear chart, particularly near the driplines and in the superheavy mass region. Its role is not limited to nuclear structure alone; it also supports interdisciplinary research connecting nuclear physics with astrophysics, atomic physics, and applications in materials science and medical physics.

Storage rings and cooler rings represent another critical component of the modern experimental ecosystem. These devices enable the confinement and repeated circulation of ion beams, allowing for extremely high-precision measurements that are difficult or impossible to achieve in single-pass experiments. One of the most important applications of storage rings is precision mass spectrometry, where the revolution frequency of stored ions is used to determine their mass-to-charge ratios with very high accuracy. Isochronous mass spectrometry, in particular, has become a powerful technique for studying short-lived exotic nuclei, enabling the measurement of nuclear masses far from stability with unprecedented precision \cite{Glorius2023StorageRingAstro,Steck2020StorageRing}.

Such measurements play a crucial role in constraining nuclear structure models, especially in regions where shell evolution and pairing effects lead to unexpected changes in binding energies. They are also essential for astrophysical applications, where nuclear masses determine reaction Q-values, beta-decay energies, and the location of waiting points in nucleosynthesis pathways such as the rapid neutron capture process. In this way, storage ring facilities directly connect microscopic nuclear properties to macroscopic astrophysical phenomena, providing critical input for models of stellar explosions and neutron star mergers \cite{Brho}.

In addition to mass measurements, storage rings enable studies of nuclear lifetimes, decay modes, and reaction cross sections under controlled conditions. The ability to store ions for extended periods also opens possibilities for internal-target experiments, where rare isotopes interact with thin gas or electron targets, allowing for precise investigations of reaction dynamics with minimal background. These capabilities make storage rings indispensable tools for exploring the properties of nuclei with extremely short lifetimes.

Complementary to accelerators and storage rings, modern light sources and neutron facilities provide essential probes of nuclear structure and reactions. Photon beams, in particular, are widely used in nuclear spectroscopy, photodisintegration studies, and the investigation of nuclear response functions. In particular, a recent photon beamline, so-called Shanghai Laser-Electron Gamma source (SLEGS) \cite{SLEGS1,SLEGS2} which 
is a compact, high-brilliance gamma-ray facility based on Compton backscattering between laser photons and relativistic electron beams at the Shanghai Synchrotron Radiation Facility. By providing quasi-monochromatic and tunable gamma rays in the MeV range, SLEGS enables precision studies of photonuclear reactions, nuclear structure, and astrophysically relevant processes \cite{SLEGS3,SLEGS4,SLEGS5,SLEGS6}. Its high intensity and energy resolution make it particularly well suited for measuring reaction cross sections near threshold, offering important inputs to nuclear astrophysics and fundamental nuclear physics. As part of China’s advanced photon science infrastructure, SLEGS also supports applications in nuclear technology, materials science, and detector development \cite{SLEGS-iso,SLEGS-pos}.

Similarly, advanced neutron sources provide unique access to  multidisciplinary research and applications, including neutron-induced reactions, which are of central importance for both nuclear technology and astrophysical nucleosynthesis. In China, 
China Spallation Neutron Source (CSNS) is a major national research facility dedicated to neutron science and multidisciplinary applications. Located in Dongguan, CSNS is operated by the Institute of High Energy Physics under the Chinese Academy of Sciences. It employs a high-power proton accelerator to bombard a heavy metal target, producing intense pulsed neutron beams through the spallation process. These neutrons are then moderated and delivered to a suite of advanced instruments for studies in condensed matter physics, materials science, chemistry, engineering, and biology. As one of the world’s leading pulsed neutron sources, CSNS provides crucial capabilities for probing the structure and dynamics of matter at the atomic scale and supports both fundamental research and industrial innovation \cite{CSNS,CSNS2}.

\begin{facilitysidebar}{Sidebar 4. Photon, neutron, and nuclear-data facilities}
\facilityimage{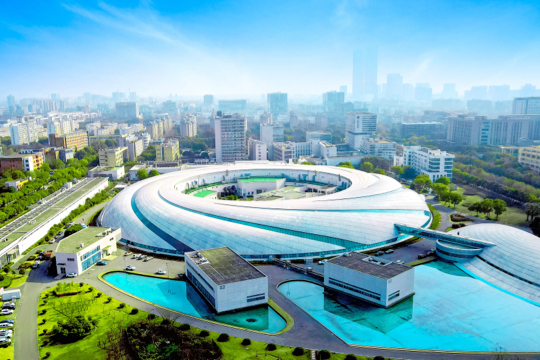}{A bird's view for Shanghai Synchrotron Radiation Facility (SSRF) inside which a beamline for Shanghai laser electron gamma source is located.}
\facilityentry{Photonuclear response}
{SLEGS at SSRF}
{Quasi-monochromatic gamma beams support photonuclear reactions, giant-resonance studies, nuclear polarizabilities, isotope-specific response functions, parity-sensitive observables, and astrophysically relevant photodisintegration measurements \cite{SLEGS1,SLEGS2,SLEGS3,SLEGS4,SLEGS5,SLEGS6,SLEGS-iso,SLEGS-pos}. The key capability is controlled photon energy and polarization combined with high-efficiency neutron, charged-particle, and gamma detection. This makes SLEGS a bridge between nuclear-structure response, astrophysical reaction rates, and photonuclear data needed in technology.}

\facilityentry{Neutron-induced reactions}
{CSNS, Back-n, and nuclear-data platforms}
{Pulsed neutron sources provide time-of-flight measurements for capture, fission, scattering, activation, threshold reactions, and detector-response studies. They are the facility class that most directly connects basic reaction physics to evaluated nuclear data, reactor design, transmutation, shielding, radiation safety, and isotope production \cite{CSNS,CSNS2}. The planning implication is that nuclear-data facilities need stable beam delivery, benchmark materials, shared detector standards, open covariance data, and close coupling to evaluated libraries rather than isolated one-off measurements.}
\end{facilitysidebar}

The experimental landscape is further enriched by a rapidly evolving generation of detector systems. Modern nuclear physics experiments increasingly rely on highly segmented, large-acceptance detector arrays capable of simultaneously measuring multiple reaction products with high efficiency and resolution. These systems are designed to capture complex final states involving charged particles, neutrons, gamma rays, and heavy fragments, often in coincidence. As a result, they enable complete kinematic reconstruction of nuclear reactions, providing detailed insight into reaction mechanisms and nuclear structure effects.

Key technological trends in detector development include increased granularity, improved energy and timing resolution, and enhanced particle identification capabilities. Silicon detectors, scintillator arrays, time projection chambers, and advanced gamma-ray tracking detectors all play important roles in this ecosystem. In particular, timing resolution at the level of tens of picoseconds is becoming increasingly important for distinguishing reaction channels and identifying short-lived intermediate states. At the same time, large solid-angle coverage ensures high detection efficiency for rare processes, which is crucial when working with low-intensity radioactive beams.

Another major transformation in modern experimental nuclear physics is the integration of detector systems with real-time data processing and online analysis frameworks. As experimental rates and data volumes continue to increase, traditional offline analysis pipelines are becoming insufficient. Instead, modern experiments are increasingly designed with embedded data acquisition and processing systems that allow for real-time event reconstruction, filtering, and selection. This integration not only improves data quality but also enables adaptive experimental strategies, where beam conditions and detector configurations can be optimized dynamically based on preliminary results.

Looking toward future frontier experiments, it is becoming clear that performance gains will not arise from isolated improvements in individual components alone. Instead, the next generation of breakthroughs will depend on facility-scale optimization, in which accelerator systems, target designs, detector architectures, and data acquisition frameworks are developed in a fully integrated manner. This holistic approach allows for the coherent optimization of beam quality, luminosity, detection efficiency, and data throughput, ensuring that the entire experimental chain operates at maximal effectiveness.

In this context, the concept of a nuclear physics facility is itself evolving \cite{Hutton2023ERL}. Rather than being viewed as a collection of independent instruments, modern facilities are increasingly designed as unified research ecosystems. In such systems, accelerators, storage rings, detector arrays, and computational infrastructure are tightly interconnected, enabling seamless transitions between different experimental modes and research programs. This integration is essential for addressing the most challenging problems in nuclear science, where rare processes, weak signals, and complex final states require maximal experimental sensitivity and flexibility.

In summary, accelerator complexes, storage rings, light sources, and detector systems form the technological foundation of modern nuclear physics. Their continued development not only expands the range of accessible experiments but also fundamentally shapes the scientific questions that can be addressed. As facilities become more powerful and more integrated, they will play an increasingly central role in advancing our understanding of nuclear structure, reactions, astrophysical processes, and the fundamental properties of strongly interacting matter.

\subsection{Precision metrology, isotopes, and cross-disciplinary technologies}

\begin{facilitysidebar}{Sidebar 5. Isotope, radiomedicine, and precision-metrology platforms}
\facilityimage{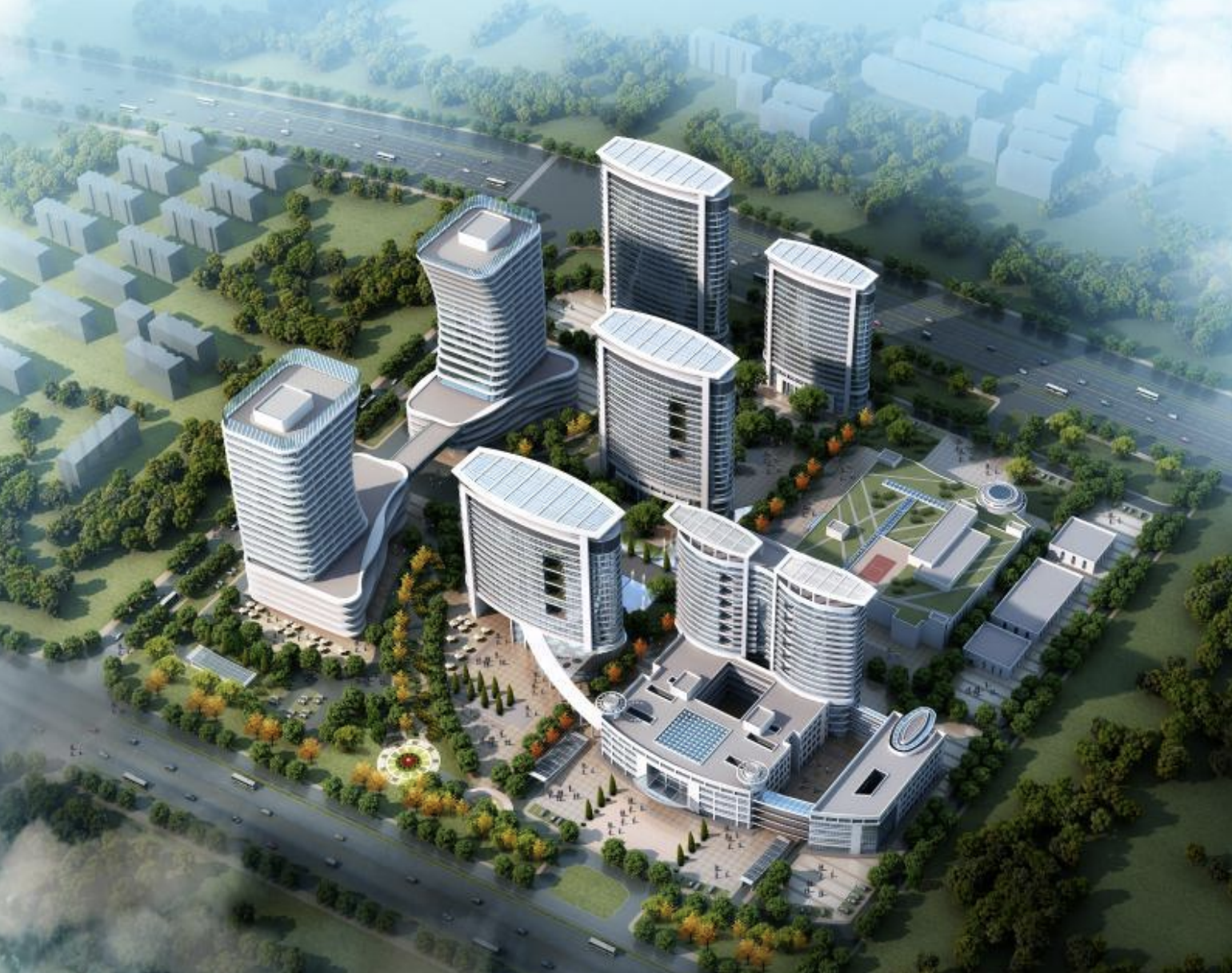}{A bird's view of Lanzhou heavy-ion tumor therapy center.}
\facilityentry{Isotopes and radiomedicine}
{Accelerator, reactor, and separation facilities}
{Medical-isotope production, theranostic radionuclides, FLASH radiotherapy, tracer science, and radiometric standards depend on reliable reaction cross sections, high-power targetry, chemical separation, dosimetry, and detector calibration. These applied platforms feed back into basic nuclear physics through improved decay data, reaction modeling, microcalorimetry, compact timing detectors, and accelerator-based beam diagnostics \cite{Pomme2022Radionuclide,Muller2024Microcalorimeter,RMP}. Their facility needs include isotope supply chains, hot-cell chemistry, quality assurance, clinical translation, and common data standards for dose and decay properties.}

\facilityentry{Nuclear clocks and quantum metrology}
{$^{229}$Th and precision spectroscopy platforms}
{The $^{229}$Th nuclear-clock program links nuclear structure, laser spectroscopy, isotope preparation, ion trapping or solid-state host control, and searches for fundamental-constant variation. It represents a precision frontier in which isotope availability, radiochemical purity, electronic-bridge effects, decay-channel control, and metrological stability become as decisive as traditional beam intensity \cite{Berengut2025IsotopeShift,Fortier2023FrequencyComb,RN115,Yamaguchi2024}. Long-range planning should therefore connect nuclear spectroscopy laboratories with atomic-clock, quantum-optics, and isotope-production infrastructure.}
\end{facilitysidebar}

One of the most dynamic features of contemporary nuclear science is its increasing integration with adjacent scientific and technological domains. While nuclear physics was historically associated primarily with nuclear structure, reactions, and energy production, it has become a broad enabling discipline with applications in precision metrology, rare-event detection, medicine, materials science, chemistry, and emerging quantum technologies~\cite{Berengut2025IsotopeShift,Fortier2023FrequencyComb,Yamaguchi2024,RN115}. This transformation is driven by two complementary developments: the improved experimental control over nuclear systems, including the production, enrichment, manipulation, and precision measurement of isotopes, and the recognition that nuclear transitions, radioactive processes, and nuclear recoil responses provide sensitive probes of physical environments and fundamental interactions that are often inaccessible to conventional atomic or molecular techniques~\cite{Pomme2022Radionuclide,Muller2024Microcalorimeter,Anand2014DMResponse,Kolos2022NuclearDataNeeds}.

In this expanded landscape, nuclear science is no longer confined to traditional reactor-based applications or accelerator-driven basic research. Instead, it now supports a wide range of interdisciplinary technologies, including deep-underground rare-event searches, nuclear clocks and precision spectroscopy, radiopharmaceutical production and therapy, advanced radiation medicine, low-background detection, nuclear-security imaging, and isotope-enabled tracing techniques in chemistry, biology, and environmental science~\cite{Bo2025PandaX4T,Abdukerim2025PandaXxT,Berengut2025IsotopeShift,Yamaguchi2024,Zhang2025Radiopharmaceuticals,Sgouros2020RPT,RMP,AlHamrashdi2019NuclearSecurity}. These applications are not peripheral to nuclear physics; rather, they are increasingly shaped by advances in nuclear theory, accelerator technology, detector systems, isotope production and enrichment, low-background measurement, and nuclear-data capabilities~\cite{Kolos2022NuclearDataNeeds,SLEGS-iso,CSNS}. As a result, nuclear science is evolving into a technological platform that connects fundamental physics with practical and societal applications.

Deep-underground searches for neutrinoless double-beta decay ($0\nu\beta\beta$) and particle dark matter provide representative examples of the enabling role of nuclear science. Their primary goals lie in neutrino physics, particle physics, and cosmology, but their experimental reach depends critically on nuclear-science platforms: isotope enrichment and purification, radiopure target preparation, low-background gamma-ray and neutron spectroscopy, time-projection chambers and cryogenic detectors, nuclear-recoil calibration, and detector-response modeling. Nuclear theory provides an equally essential link, since nuclear matrix elements connect a possible $0\nu\beta\beta$ signal to the underlying lepton-number-violating mechanism, while nuclear form factors and response functions enter the interpretation of dark-matter scattering data \cite{Agostini2023NuDBD,Schumann2019DirectDetection,Anand2014DMResponse,Yao2022NME,Belley2024NME}.

China's rapidly developing underground infrastructure has become an important part of this effort. The China Jinping Underground Laboratory (CJPL), with its exceptional rock overburden and low cosmic-ray background, provides a shared environment for rare-event searches and low-background technology development \cite{Cheng2017CJPL}. At CJPL, CDEX employs low-threshold point-contact germanium detectors for dark-matter searches and $^{76}$Ge double-beta-decay studies, whereas PandaX uses large liquid-xenon time-projection chambers for dark-matter detection and searches involving $^{136}$Xe \cite{Jiang2018CDEX10,Zhang2024CDEX0nu,Bo2025PandaX4T,Zhang2025PandaX0nu}. Dedicated concepts such as PandaX-III and N$\nu$DEx-100 further extend this program through high-pressure gas time-projection chambers with enriched $^{136}$Xe and $^{82}$SeF$_6$, respectively \cite{Chen2017PandaXIII,Cao2024NvDEx}.

Neither class of experiment has yet yielded a confirmed laboratory signal, and future sensitivity will depend on more than increasing target mass and exposure. Radioactive impurities, cosmogenic activation, radon, neutron-induced backgrounds, recoil-response uncertainties, and ultimately neutrino backgrounds will impose increasingly stringent requirements. Continued advances in isotope separation, radiochemical purification, material assay, neutron calibration, quenching and ionization-yield measurements, evaluated nuclear data, and uncertainty-controlled nuclear-structure calculations will therefore remain essential. The proposed PandaX-xT observatory illustrates how future multi-tonne platforms may combine dark-matter, neutrino, and double-beta-decay measurements within a common nuclear-science and low-background technology infrastructure \cite{Abdukerim2025PandaXxT}. These rare-event programs thus form a natural bridge from underground low-background technology to the broader precision frontier, where nuclear systems themselves can serve as exceptionally sensitive probes of fundamental interactions.

\begin{figure}[!htb]
    \includegraphics[width =1\linewidth]{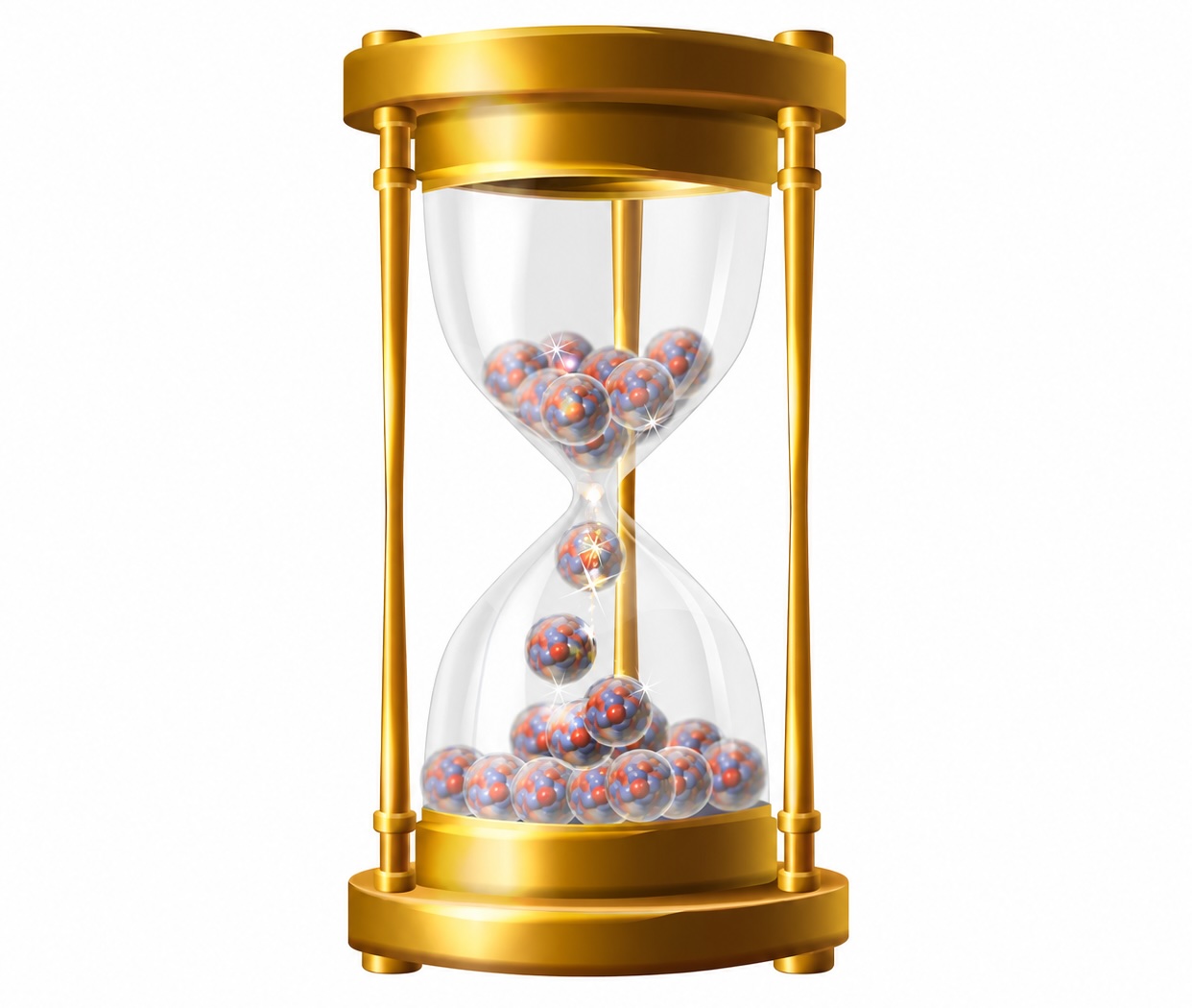}
    \caption{Cartoon of $^{229}$Th clock. {Original schematic prepared by the author.}}
    \label{Th229_clock}
\end{figure}

A more recent and particularly striking example of this convergence is the emerging concept of the nuclear clock based on the low-lying isomeric transition in $^{229}$Th (see Fig.~\ref{Th229_clock}). Unlike conventional atomic clocks, which rely on electronic transitions in atomic shells, a nuclear clock exploits a transition within the nucleus itself. The exceptionally low excitation energy of the $^{229}$Th isomer makes it uniquely accessible to laser-based spectroscopy, opening the possibility of a clock that is significantly less sensitive to external electromagnetic perturbations than existing atomic standards. This property has profound implications for precision metrology, as it could enable timekeeping systems with unprecedented stability and accuracy.

The development of a nuclear clock is not only a technical achievement in spectroscopy but also a fundamental test of nuclear structure theory and quantum electrodynamics in a novel regime. The precise energy, lifetime, and decay properties of the $^{229}$Th isomer remain active areas of experimental and theoretical investigation. Small uncertainties in these quantities directly affect the feasibility and performance of the proposed clock. Moreover, because nuclear energy levels are sensitive to fundamental interactions within the nucleus, such a clock could potentially be used to probe temporal variations of fundamental constants, search for physics beyond the Standard Model, and explore possible couplings between nuclear states and external fields such as gravitational or dark matter backgrounds. In this sense, the nuclear clock represents a unique intersection of nuclear physics, atomic physics, quantum technology, and fundamental metrology \cite{RN115,Yamaguchi2024}.

Beyond precision timekeeping, nuclear science plays a crucial role in medical applications, particularly through the production and utilization of radioisotopes. Nuclear medicine relies on the ability to produce isotopes with specific decay properties that can be used for both diagnostic imaging and therapeutic purposes. Techniques such as positron emission tomography (PET) and single-photon emission computed tomography (SPECT) depend on carefully selected radionuclides that emit detectable radiation while maintaining suitable half-lives for biological applications. The production of these isotopes requires advanced accelerator systems, reactor facilities, and separation technologies, all of which are rooted in nuclear physics research. {China's nuclear-medicine sector now includes nearly 1200 hospitals and approximately 13,000 professionals, serving more than 3.9 million patients annually~\cite{Yang2024ChinaNuclearMedicine}. A resilient domestic supply chain must combine reactor- and accelerator-based production of strategic isotopes---including $^{99}$Mo, $^{177}$Lu, $^{64}$Cu, $^{89}$Zr, and $^{225}$Ac---with targetry, enriched feedstocks, radiochemical separation, activity standards, good-manufacturing-practice quality control, and reliable distribution.}

In recent years, radiotherapy has undergone significant transformation due to advances in beam delivery and radiation biology. One of the most notable developments is FLASH radiotherapy, which delivers ultra-high dose rates of radiation in extremely short time intervals. Experimental studies have shown that FLASH irradiation can reduce damage to healthy tissue while maintaining or enhancing tumor control, although the underlying biological mechanisms are still under active investigation. This phenomenon has stimulated extensive interdisciplinary research involving nuclear physics, medical physics, radiobiology, and clinical oncology. It also highlights the importance of precise control over beam characteristics, including intensity, timing structure, and energy deposition profiles, all of which depend critically on accelerator technology and beam diagnostics \cite{RMP}. {Domestic carbon-ion therapy systems provide another representative translation chain, linking accelerator and beam-delivery physics to treatment planning, radiobiology, medical-device certification, and routine hospital operation; the Wuwei and Lanzhou centers illustrate the transition from laboratory capability to clinical service.}

The broader field of isotope-enabled science extends well beyond medicine. Isotopic labeling and radionuclide tracing provide sensitive tools for following transport processes, reaction pathways, and long-time-scale transformations in complex physical, chemical, biological, and environmental systems. These applications rely on the availability of high-purity isotopes, accurate activity standards, and well-characterized decay data, all of which require precise control over isotope production, target design, separation methods, and nuclear reaction modeling~\cite{Pomme2022Radionuclide,Muller2024Microcalorimeter,Kolos2022NuclearDataNeeds,SLEGS-iso}. In this sense, applied isotope science remains closely tied to the same nuclear-data infrastructure that supports fundamental nuclear physics.

An important feature of these cross-disciplinary applications is that they increasingly feed back into fundamental nuclear physics. The need for medically relevant isotopes has driven improvements in reaction modeling and photonuclear cross-section measurements, while precision radionuclide metrology has motivated more accurate determinations of decay schemes, branching ratios, and half-lives~\cite{SLEGS-iso,Pomme2022Radionuclide,Muller2024Microcalorimeter}. Similarly, the development of low-background rare-event searches and advanced imaging techniques has led to new detector technologies, radiopurity-control methods, calibration procedures, and analysis strategies that are now being applied in basic nuclear experiments~\cite{Chen2017PandaXIII,Bo2025PandaX4T,Abdukerim2025PandaXxT,Miernik2007b}. This bidirectional flow of knowledge illustrates the deeply interconnected nature of modern nuclear science.

Another key aspect of this interdisciplinary expansion is the role of accelerator technology as a shared infrastructure. Modern accelerator complexes are no longer dedicated solely to fundamental nuclear research; they are increasingly designed as multipurpose facilities capable of supporting isotope production, nuclear-data measurements, detector calibration, materials irradiation, neutron-response studies, and other applied programs~\cite{Zhou2022,HIAF,SLEGS1,Wang2022,CSNS,CSNS2}. The flexibility and versatility of these facilities make them central hubs for both scientific discovery and technological innovation.

At the same time, advances in detector systems, data acquisition, and real-time processing are enabling increasingly sophisticated measurements across these diverse applications. High-resolution imaging detectors, fast timing systems, cryogenic sensors, time-projection chambers, and compact radiation detectors are now used in both fundamental and applied settings~\cite{Miernik2007b,Muller2024Microcalorimeter,Chen2017PandaXIII,AlHamrashdi2019NuclearSecurity}. The integration of machine learning and data-driven analysis techniques further enhances the ability to extract meaningful information from complex experimental datasets, whether in rare-event searches, nuclear spectroscopy, medical imaging, or materials diagnostics~\cite{Boehnlein2022,MaSCPMA23,MaNST23}.

In summary, precision metrology, isotope science, rare-event detection, and cross-disciplinary technologies represent a rapidly expanding frontier of nuclear science. Far from being isolated applications, these areas are deeply rooted in advances in nuclear structure, reaction and decay theory, accelerator physics, isotope technology, low-background instrumentation, and detector science. The development of neutrinoless double-beta-decay and dark-matter experiments, nuclear clocks, medical-isotope systems, and advanced radiation therapies illustrates how nuclear physics is increasingly becoming a foundational enabling science whose impact extends well beyond its traditional disciplinary boundaries. As these fields continue to evolve, the synergy between fundamental research, shared scientific infrastructure, and applied technology will play an increasingly important role in shaping both scientific discovery and practical innovation.

\subsection{Digitalization, AI-assisted workflows, and intelligent platforms}

\begin{figure*}[htb]
\caption{Schematic relationships between the topics of nuclear physics~\cite{Boehnlein2022}. The diagram emphasizes the close connections between theory, computations (both computational science and data science as well as many elements from computer science) and experiments.}\label{fig:machine_learning}
{\includegraphics[width=1\textwidth]{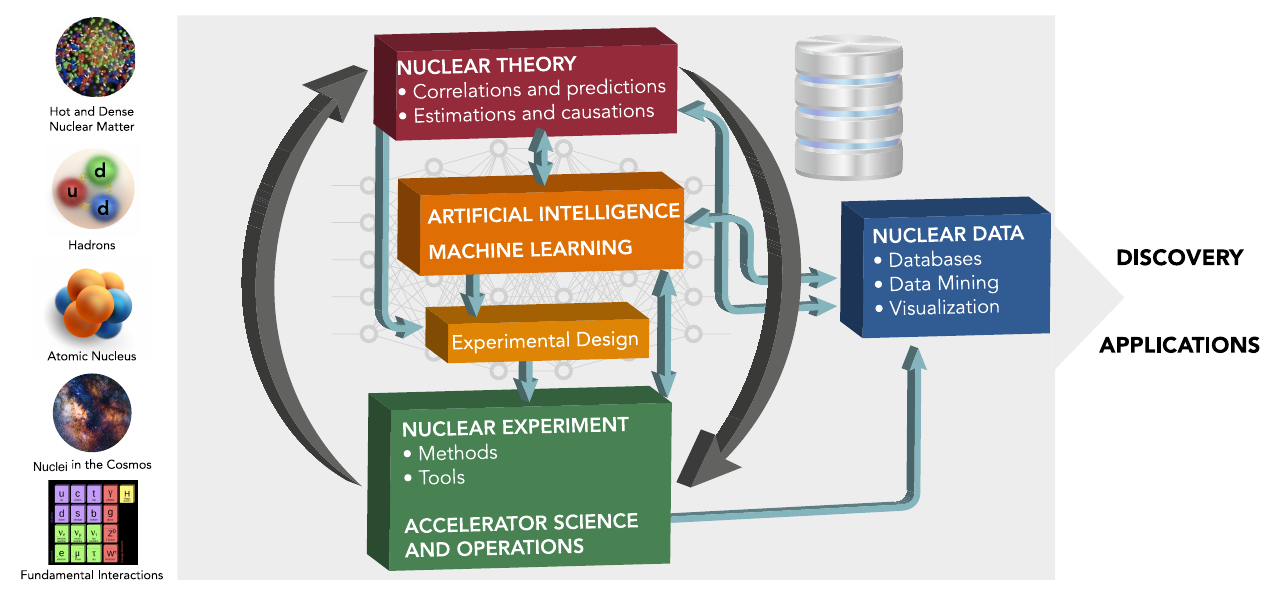}}
\end{figure*}

The ninth frontier question explicitly highlights AI-enabled experimental platforms as a central enabling capability for future nuclear science. This emphasis is well justified by both the increasing complexity of modern experiments and the rapid growth in computational demands across theoretical and data-analysis pipelines. In contemporary nuclear physics, the scale and heterogeneity of data generated by large detector arrays, high-intensity accelerators, and multi-stage simulation frameworks are reaching levels that fundamentally challenge traditional offline analysis paradigms. As a result, digitalization and artificial intelligence (AI) are no longer auxiliary tools, but are becoming integral components of the scientific workflow itself \cite{AI-Book,Jiao2024AIPhysics,AINuclearTech2025,Mondal2024DigitalTwin,Allaire2024AI4EIC} (see Fig.~\ref{fig:machine_learning}).

This transformation is driven by several converging trends. First, modern nuclear experiments often involve event rates that are too high to store or process fully using conventional approaches, necessitating real-time filtering, compression, and feature extraction directly at the data acquisition stage. Second, the complexity of observables—ranging from multi-particle correlations in heavy-ion collisions to rare decay signatures in exotic nuclei—requires sophisticated pattern recognition methods that can efficiently disentangle signal from background. Third, theoretical models, particularly in nuclear many-body physics and reaction theory, are becoming increasingly computationally expensive, often requiring large-scale simulations that span high-dimensional parameter spaces. These factors collectively motivate the development of intelligent, adaptive, and automated workflows.

Within this context, machine learning techniques, Bayesian inference methods, surrogate modeling, and automated optimization algorithms are emerging as essential components of the nuclear science toolkit \cite{MaNST23,MaSCPMA23,MaCPL23}. Machine learning, in particular, has demonstrated significant potential in classification tasks, regression problems, and anomaly detection in experimental data. For example, deep neural networks can be trained to identify rare event signatures in detector outputs, significantly improving signal extraction efficiency compared to traditional cut-based methods. Similarly, Bayesian approaches provide a principled framework for incorporating prior knowledge and systematically quantifying uncertainties, which is particularly important in a field where experimental constraints are often indirect and statistically limited.

Surrogate modeling represents another critical development, especially in theoretical nuclear physics. Many-body calculations, including those based on \textit{ab initio} methods, nuclear density functional theory, or transport simulations of heavy-ion collisions, can be computationally prohibitive when explored across large parameter spaces. Surrogate models, including Gaussian-process emulators, reduced-order models, and neural-network-based emulators, provide efficient approximations to these expensive calculations, enabling rapid parameter exploration, statistical calibration, and uncertainty quantification~\cite{Boehnlein2022,MaSCPMA23,MaNST23,Cook2025}. In nuclear astrophysics, related emulator strategies can accelerate sensitivity studies of nucleosynthesis pathways and help connect microscopic nuclear inputs with macroscopic astrophysical observables~\cite{Boehnlein2022,Kolos2022NuclearDataNeeds}.

Automated optimization techniques are also playing an increasingly important role in experimental design and accelerator operation. Modern accelerator facilities involve a large number of tunable parameters, including beam energy, intensity, focusing conditions, and target configurations. Optimizing these parameters for specific experimental goals is a high-dimensional and nonlinear problem. Machine-learning-based optimization algorithms, including Bayesian optimization, reinforcement learning, and evolutionary strategies, are being explored to improve beam stability, maximize luminosity, and enhance experimental efficiency~\cite{Boehnlein2022,Edelen2024AcceleratorML}. These approaches enable adaptive control systems that can respond dynamically to changing experimental conditions, thereby improving both data quality and operational efficiency.

However, despite these promising developments, the role of AI in nuclear science must be carefully contextualized and physically grounded. The effectiveness of data-driven methods depends critically on their integration with established physical principles, detector response models, and rigorous uncertainty quantification~\cite{Boehnlein2022,MaSCPMA23,MaNST23}. Purely data-driven approaches, without incorporation of physical constraints, risk producing results that are difficult to interpret or generalize beyond the training domain. In nuclear physics, where extrapolation to unmeasured regimes is often essential, maintaining physical consistency is particularly crucial.

One important requirement for AI-assisted workflows is therefore the incorporation of physical priors. These priors may take the form of conservation laws, symmetries, known interaction structures, or constraints derived from effective field theories. By embedding such information into machine learning architectures, one can improve both robustness and interpretability. Physics-informed and hybrid modeling approaches, which combine first-principles calculations with data-driven corrections, represent promising directions in this regard~\cite{Boehnlein2022,Furnstahl2015,Melendez2019}.

Equally important is the integration of detector response modeling into AI-based analysis pipelines. In nuclear experiments, raw signals are always filtered through complex detection systems, which introduce inefficiencies, resolution effects, and systematic biases. Accurate interpretation of experimental data therefore requires a detailed understanding of detector response functions. AI methods can be used not only to analyze data, but also to model and invert detector effects, enabling more precise reconstruction of underlying physical observables. This is particularly relevant in high-multiplicity environments such as heavy-ion collisions, where event reconstruction is intrinsically complex~\cite{Boehnlein2022,MaNST23}.

Another critical aspect is uncertainty quantification. In both experimental and theoretical nuclear physics, reliable estimates of uncertainties are essential for meaningful comparisons between theory and experiment. AI-assisted methods must therefore be designed to propagate uncertainties consistently through all stages of analysis, from raw data processing to final physical inference. Bayesian neural networks, ensemble methods, Gaussian-process emulators, and probabilistic model-comparison techniques are among the approaches being actively explored to address this challenge~\cite{Hoeting1999,Furnstahl2015,Melendez2019,Neufcourt2020b,Boehnlein2022}. The goal is not only to improve predictive accuracy, but also to provide transparent and quantifiable measures of confidence in the results.

In this broader framework, the most promising applications of AI in nuclear science can be identified in several key areas. Online event classification and trigger systems can significantly enhance the efficiency of data collection by selecting relevant events in real time. Adaptive beam tuning and accelerator optimization can improve experimental performance and stability. Inverse problems in imaging and spectroscopy can benefit from AI-based reconstruction techniques that extract physical information from incomplete or noisy data. Emulator-assisted many-body calculations can accelerate theoretical predictions and enable large-scale parameter studies \cite{Cook2025}. Finally, accelerated uncertainty propagation in astrophysical reaction networks and reactor simulations can improve the reliability of macroscopic models based on microscopic nuclear inputs. {Accordingly, AI should be treated as a cross-cutting platform rather than only a data-analysis tool: priority applications include reactor digital twins, autonomous experiments, evaluated nuclear-data workflows, patient-specific dose optimization, predictive maintenance, equipment-health management, and anomaly-aware safety monitoring.}

Crucially, the ultimate value of intelligent platforms in nuclear science lies not in isolated algorithmic performance, but in their ability to enhance the end-to-end scientific cycle. This includes data acquisition, preprocessing, analysis, theoretical modeling, and physical interpretation. When properly integrated, AI-driven workflows can shorten the feedback loop between experiment and theory, enabling more efficient hypothesis testing and model refinement. This closed-loop approach is particularly powerful in a field where experiments are costly, data are complex, and theoretical models are computationally intensive.

In conclusion, digitalization and AI-assisted methodologies are rapidly becoming foundational elements of modern nuclear science. Their impact will be determined not simply by technological sophistication, but by their integration with physical principles, experimental realities, and theoretical consistency. When these conditions are met, intelligent platforms have the potential to fundamentally transform how nuclear science is conducted, enabling a more adaptive, efficient, and insight-driven research ecosystem that spans from fundamental nuclear structure to astrophysical applications.

\section{Translational and strategic frontiers}
\subsection{Advanced fission and fusion for sustainable energy}

Energy remains one of the most visible and societally consequential interfaces of nuclear science. Unlike many other frontiers in nuclear physics, the energy frontier directly couples fundamental nuclear processes to large-scale engineering systems, economic constraints, environmental considerations, and long-term sustainability goals. However, the sixth frontier question should not be interpreted narrowly as a reactor engineering problem. Instead, it represents a deeply interdisciplinary physics challenge that spans neutron transport, many-body nuclear data, radiation–matter interaction, extreme-condition materials science, plasma physics, and system-level control theory. In this sense, advanced nuclear energy development is as much a problem in complex physical systems as it is in technological optimization.

\begin{figure}[!htb]
    \includegraphics[width =1\linewidth]{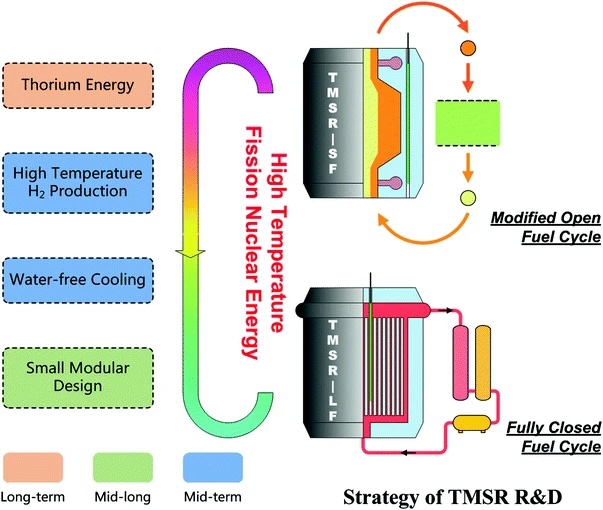}
    \caption{The thorium molten-salt reactor \cite{DAI2017531} program is a representative system-level platform linking reactor physics, materials, chemistry, and fuel-cycle strategy.}
    \label{NucPower}
\end{figure}

Both advanced fission \cite{Abram2008GenerationIV} and controlled fusion \cite{Meschini2023FusionCommercial}  require sustained and coordinated progress across multiple scientific layers. At the most fundamental level, reliable nuclear data—including cross sections, decay channels, fission fragment distributions, and neutron emission spectra—form the backbone of predictive reactor and plasma modeling. These inputs are not merely auxiliary parameters; they determine the stability, efficiency, and safety margins of entire energy systems (Fig.~\ref{NucPower}). At the intermediate scale, accurate transport modeling of neutrons, photons, and charged particles is essential for describing energy deposition, breeding processes, and shielding requirements. At the macroscopic scale, structural materials must withstand extreme irradiation environments characterized by high neutron fluxes, displacement damage, helium/hydrogen production, and long-term degradation effects. Finally, at the system level, advanced control strategies and diagnostics are required to ensure stable operation under dynamic and often nonlinear feedback conditions.

In advanced fission systems, the primary scientific and technological goals include enhanced passive safety, improved fuel utilization, reduced production of long-lived radioactive waste, and increased flexibility in deployment and fuel cycle design. These objectives are closely interrelated. For example, improving fuel utilization often requires higher burnup levels, which in turn intensify material degradation and require more robust cladding and structural materials. Similarly, reducing long-lived waste is closely connected to advances in transmutation technologies and fuel reprocessing strategies, which depend sensitively on accurate neutron-induced reaction data and decay heat modeling. Fast neutron reactors, molten salt systems \cite{TMSR,TMSR2,NucPower}, and accelerator-driven subcritical systems \cite{CIADS,CIADS2} represent different technological pathways toward these goals, each with distinct physics challenges and engineering constraints.

{China's recent reactor portfolio illustrates the progression from deployment to advanced concepts. The scaled construction and commercial operation of Hualong One represent an engineering-scale contribution to Generation-III pressurized-water-reactor design, supply-chain localization, and low-carbon power infrastructure~\cite{CNNC2021Hualong}. The Shidaowan HTR-PM entered commercial operation in December 2023, demonstrating modular high-temperature gas-cooled reactor technology and its potential for process heat~\cite{IAEA2024SMR}. The 125-MWe ACP100/Linglong One, now under construction, extends this portfolio toward modular multipurpose systems for electricity, heat, industrial steam, and desalination~\cite{IAEA2023ACP100}. Fast-reactor development, including CFR-600-related construction and commissioning activity, should be discussed cautiously as a demonstration pathway coupled to MOX fuel, sodium safety, transmutation, and closure of the fuel cycle rather than as established commercial maturity~\cite{IAEA2024FastReactors}. Likewise, thorium molten-salt development is not simply a choice of reactor type: it requires integrated solutions for online fuel chemistry, corrosion, tritium control, source-term behavior, licensing, and waste streams.}

For instance, China’s Thorium Molten Salt Reactor (TMSR) experimental program is a pioneering effort to develop next-generation nuclear energy systems with enhanced safety, efficiency, and sustainability. Led by the Shanghai Institute of Applied Physics under the Chinese Academy of Sciences, the TMSR project explores the use of liquid fuel and thorium-based fuel cycles to achieve intrinsic safety and low-pressure operation. The experimental reactor is designed to demonstrate key technologies such as online fuel reprocessing, corrosion-resistant materials, and efficient heat utilization. By advancing thorium utilization and molten salt technology, TMSR aims to provide a promising pathway toward clean energy with reduced long-lived radioactive waste and improved resource efficiency. In addition, China’s Accelerator-Driven Subcritical System (CIADS) is a flagship project aimed at developing advanced nuclear energy technologies with enhanced safety and sustainability. Led by the Chinese Academy of Sciences, CIADS combines a high-power proton accelerator with a subcritical reactor core, where spallation neutrons drive the fission process in a controlled manner. This inherently safe design eliminates the risk of criticality accidents while enabling efficient transmutation of long-lived nuclear waste. The CIADS program also serves as a multidisciplinary platform for nuclear physics, materials science, and reactor engineering, contributing to the long-term development of clean nuclear energy and closed fuel cycle strategies.

From a nuclear physics perspective, one of the central challenges in advanced fission is the accurate description of neutron-induced fission processes across a wide range of isotopes and energies. This includes not only actinides relevant to energy production but also minor actinides and fission products that contribute to long-term radiotoxicity. Theoretical models of fission dynamics, including macroscopic–microscopic approaches, density functional theory, and stochastic Langevin-type descriptions, are continuously being refined to better reproduce experimental fission barriers, fragment mass distributions, and prompt neutron emission characteristics. However, significant uncertainties remain, particularly in extrapolations to nuclei far from stability or to energy regimes not yet experimentally accessible.

In parallel, fusion energy research faces a distinct but equally complex set of challenges \cite{Hesch2024FusionProgress}. The central goal is the achievement of sustained burning plasmas with sufficient energy gain to enable net energy production \cite{ICF}. This requires precise control of plasma confinement, stability, and energy balance in regimes characterized by extreme temperature, strong magnetic fields, and highly nonlinear collective behavior. Magnetic confinement fusion, as realized in Tokamak and stellarator configurations, relies on maintaining plasma equilibrium over long timescales while suppressing instabilities such as edge-localized modes and disruptions. {China's magnetic-confinement program combines plasma-performance milestones with reactor-oriented engineering. EAST maintained steady-state high-confinement plasma operation for 1066~s in 2025, while the Comprehensive Research Facility for Fusion Technology (CRAFT) is validating major subsystems and components required for future fusion reactors~\cite{CAS2025EAST,CAS2025CRAFT}. CFETR is being developed as a bridge from experimental devices toward an engineering test reactor. Together with China's participation in ITER, these programs mark a shift from plasma-physics demonstration toward integrated fusion nuclear engineering.}

Inertial confinement fusion, on the other hand, depends on the rapid compression and heating of fuel pellets to achieve ignition conditions within extremely short timescales \cite{ICF2}.
Recent experimental advances in inertial confinement fusion have provided important evidence supporting the feasibility of achieving ignition and energy gain under controlled laboratory conditions. These results have significantly strengthened the broader scientific credibility of fusion as a long-term energy option \cite{PhysRevLett.135.035101}. At the same time, they have highlighted the importance of precision control over laser-plasma interactions, symmetry of compression, and energy coupling efficiency. Magnetic fusion research, meanwhile, continues to make steady progress in confinement performance, turbulence suppression, and integrated plasma operation scenarios, although major challenges remain in scaling toward reactor-relevant conditions.

\begin{facilitysidebar}{{Sidebar 6. Nuclear-energy and fuel-cycle test platforms}}
\facilityimage{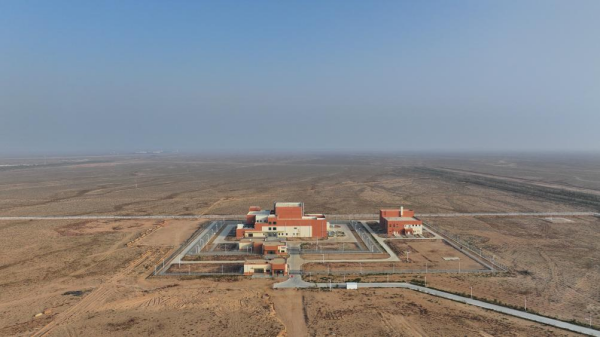}{Aerial view of the building housing the 2 MW liquid-fueled thorium-based molten salt experimental reactor \cite{NucPower} in Minqin County, Gansu Province, China.}
\facilityentry{Advanced fission and transmutation}
{TMSR and CiADS}
{Thorium molten-salt reactors explore liquid-fuel operation, thorium utilization, online fuel chemistry, corrosion-resistant materials, passive safety, tritium control, and coupled neutronics--thermal-hydraulics \cite{TMSR,TMSR2,NucPower}. CiADS connects high-power accelerators, spallation neutron production, subcritical reactor physics, beam-trip reliability, and transmutation of long-lived radionuclides \cite{CIADS,CIADS2}. Together they require nuclear data, materials irradiation, chemistry, control systems, and safety analysis to be designed as one integrated facility ecosystem.}

\facilityentry{Fusion and irradiation materials}
{EAST, CFETR, ITER-linked programs, and irradiation test needs}
{Magnetic and inertial fusion programs require burning-plasma control, plasma-facing materials, high-heat-flux components, tritium breeding, disruption mitigation, and diagnostics under extreme neutron and electromagnetic environments. Their development depends on the same nuclear-data, materials-damage, AI-control, and safety-analysis infrastructure used in advanced fission \cite{Meschini2023FusionCommercial,Hesch2024FusionProgress,ICF,ICF2,PhysRevLett.135.035101}. A facility roadmap should therefore couple plasma performance milestones with irradiation testing, nuclear diagnostics, and materials qualification.}

\facilityentry{Fuel-cycle stewardship}
{Partitioning, transmutation, and waste-governance infrastructure}
{The long-term fuel cycle links decay data, neutron-induced reactions, actinide chemistry, fuel behavior, repository science, safeguards, emergency response, and public governance. It therefore sits at the interface between facility physics and institutional long-range planning \cite{IAEA2025FuelCycleRev2,Salvatores2011PNT,NEA2024SMRWasteWorkshop}. The relevant infrastructure includes hot laboratories, underground disposal research, evaluated data centers, transmutation test systems, and long-duration monitoring platforms.}
\end{facilitysidebar}

A particularly critical aspect of both fission and fusion systems is the behavior of structural and functional materials under extreme irradiation environments \cite{Peluso2023FusionChallenges}. In fission reactors, materials must withstand sustained neutron bombardment that leads to atomic displacements, defect accumulation, swelling, embrittlement, and phase instability. In fusion reactors, the situation is further complicated by the presence of high-energy 14 MeV neutrons, which produce significant transmutation effects and helium accumulation. {This statement refers specifically to the deuterium-tritium fuel cycle. Alternative fuels such as deuterium-$^3$He would reduce the primary 14-MeV-neutron burden, but require substantially higher plasma performance and still generate secondary neutrons through competing reactions. Extraterrestrial $^3$He resources, including lunar inventories, are therefore a long-term possibility rather than a near-term solution; credible assessment must include extraction, transport, breeding alternatives, and the full energy balance.} Developing materials with sufficient radiation tolerance, thermal stability, and mechanical strength under these conditions remains one of the most important bottlenecks in realizing practical nuclear energy systems.

Another key dimension is the development of advanced diagnostics and digital control systems. Modern nuclear energy systems increasingly rely on real-time monitoring of plasma behavior, neutron flux distributions, temperature fields, and structural integrity. High-resolution diagnostics, combined with fast data acquisition and AI-assisted control algorithms, enable more precise and adaptive operation. This is particularly important for fusion plasmas, where small perturbations can lead to large-scale instabilities, and for advanced fission systems, where safety margins must be continuously monitored and maintained under varying operational conditions.

From a broader strategic perspective, both advanced fission and fusion are now understood as integral components of long-term energy planning. Long-range assessments emphasize that nuclear energy development cannot be separated from broader societal objectives, including decarbonization, energy security, infrastructure resilience, and technological sovereignty \cite{nsac2023lrp,nupecc2024lrp,Abram2008GenerationIV,IAEA2025FuelCycleRev2}. In this context, nuclear energy is increasingly viewed not as an isolated technological option, but as part of a diversified and interconnected energy ecosystem.

A productive way to conceptualize the energy frontier is therefore as a convergence problem across multiple disciplines. Progress depends on the integration of fundamental nuclear physics, plasma physics, materials science, computational modeling, digital control systems, and safety engineering. Importantly, these components are not independent modules that can be optimized separately; rather, they form a tightly coupled system in which improvements in one area often depend on advances in others. For example, improved nuclear data directly enhances reactor simulations, which in turn inform materials design and safety analysis. Similarly, advances in AI-based control systems can improve plasma stability, which feeds back into engineering design constraints.

Looking forward, it is increasingly clear that the next decisive advances in nuclear energy will arise not from isolated breakthroughs in individual technologies, but from integrated system-level approaches. Such approaches emphasize co-design principles, in which physics modeling, engineering development, and control strategies are developed simultaneously rather than sequentially. This integrated perspective is essential for addressing the complexity of real-world nuclear energy systems and for ensuring that they meet the stringent requirements of safety, efficiency, and sustainability.

In summary, advanced fission and fusion energy represent a deeply interconnected frontier of nuclear science, where fundamental physics and large-scale engineering converge. Their successful development will depend on sustained interdisciplinary collaboration and a system-level understanding of how microscopic nuclear processes propagate through complex macroscopic infrastructures to produce controlled and sustainable energy.

\subsection{Fuel cycle, waste management, and long-term nuclear governance}

\begin{figure*}[htb]
\floatbox[{\capbeside\thisfloatsetup{capbesideposition={right,top},capbesidewidth=0.2\textwidth}}]{figure}[\FBwidth]
{\caption{
    Conceptual framework for fuel cycle sustainability, radioactive waste management, and long-term nuclear governance. The diagram highlights six interdependent pillars—nuclear data, fuel behavior, partitioning and transmutation, advanced reactors/accelerator-driven systems, geological disposal, and governance and public trust—that together support technically credible, environmentally responsible, and socially durable nuclear energy systems. {Original schematic prepared by the author.}
    }\label{fig:Nuclear_fuel}}
{\includegraphics[width=0.75\textwidth]{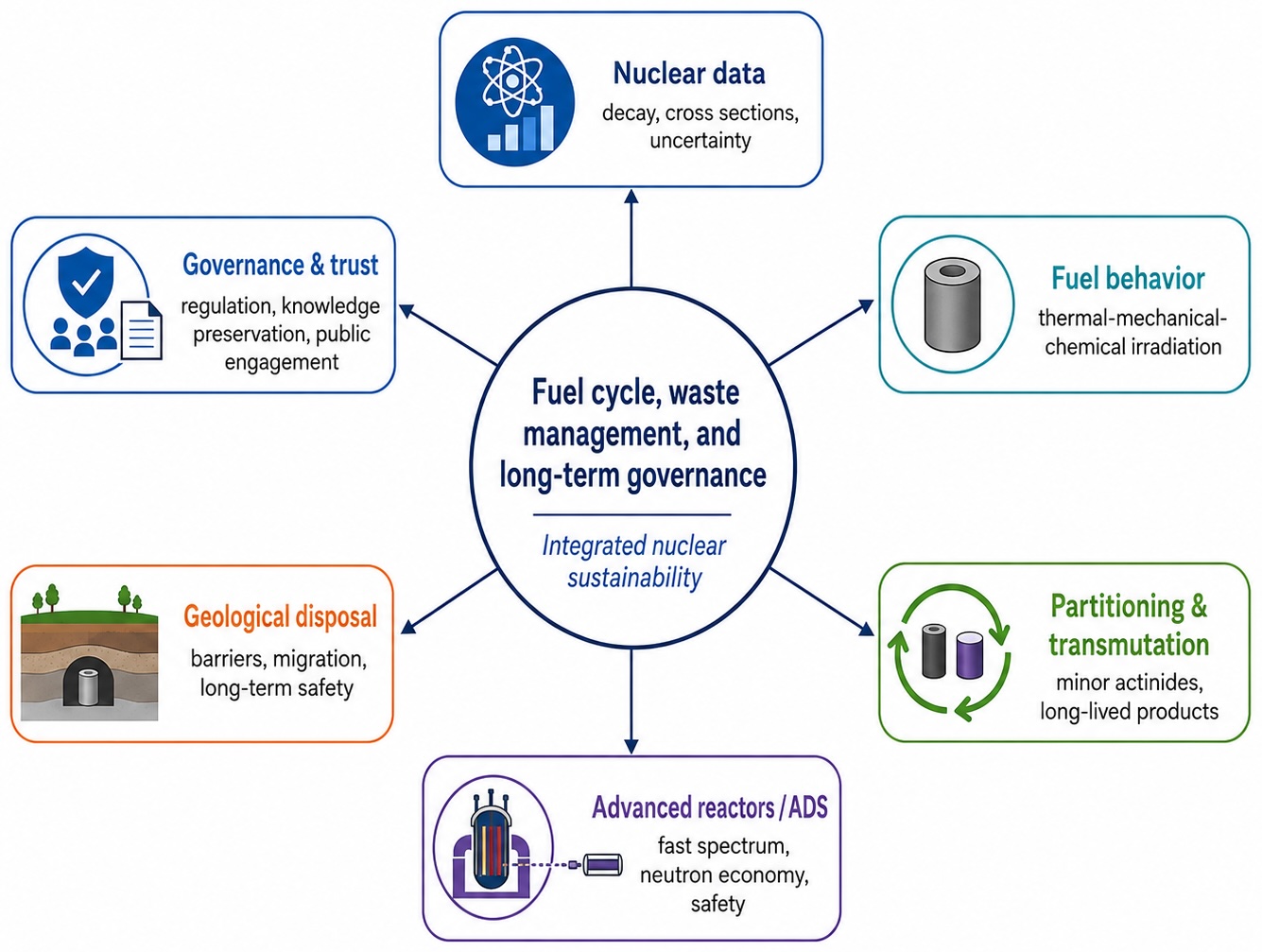}}
\end{figure*}

The long-term sustainability of nuclear energy systems depends not only on efficient and safe power generation, but also on the integrity of the entire fuel cycle, the management of high-level radioactive waste, and the resilience of systems under both operational and accidental conditions \cite{IAEA2025FuelCycleRev2,Salvatores2011PNT}. Unlike many other areas of energy technology, nuclear fuel cycles inherently involve long-lived radioactive materials, multi-decadal or even multi-millennial time horizons, and complex couplings between physics, chemistry, materials science, engineering design, and institutional governance. As a result, this domain cannot be treated as a purely technical add-on to reactor physics; it must instead be understood as an integral component of the nuclear enterprise, with its own set of fundamental scientific challenges and strategic implications (see Fig.~\ref{fig:Nuclear_fuel}).

From a scientific perspective, a central requirement is the availability of accurate and comprehensive nuclear data relevant to decay chains, neutron-induced reactions, and transmutation processes. The evolution of spent nuclear fuel is governed by a large network of coupled nuclear reactions and radioactive decays, involving actinides, fission products, and activation products. Predicting the time-dependent radiotoxicity, heat load, and neutron emission of spent fuel assemblies requires not only precise decay data, but also reliable cross sections for neutron capture, fission, and spallation reactions over a broad energy range. Uncertainties in these quantities directly propagate into uncertainties in fuel cycle analysis, repository design, and transmutation performance, making nuclear data evaluation a foundational component of the field. {The corresponding national ecosystem should integrate new measurements from facilities such as SLEGS and CSNS/Back-n with evaluated libraries, covariance files, benchmark experiments, detector-response models, versioned open formats, and reproducible processing workflows. This end-to-end structure is essential if nuclear data are to remain traceable from raw measurement to reactor, shielding, isotope-production, and astrophysical applications.}

Equally important is the development of improved multiphysics simulation frameworks for fuel behavior under irradiation. Nuclear fuel is a complex, evolving material system in which thermal, mechanical, chemical, and irradiation-induced processes are strongly coupled. Under reactor conditions, fuel undergoes significant microstructural changes, including fission gas bubble formation, swelling, cracking, phase transformations, and redistribution of chemical species. These effects influence thermal conductivity, mechanical stability, and overall fuel performance. Accurate modeling of these phenomena requires the integration of atomistic simulations, mesoscale microstructure evolution models, and continuum-level fuel performance codes. The multiscale nature of the problem makes it one of the most challenging areas in computational nuclear engineering.

In parallel, partitioning and transmutation represent an important scientific and technological frontier in radioactive-waste management. Partitioning refers to the chemical separation of long-lived radionuclides, particularly minor actinides and selected long-lived fission products, from spent nuclear fuel. Transmutation then seeks to convert these nuclides into shorter-lived or stable species through neutron irradiation or other nuclear processes. In principle, this strategy offers a pathway to reduce the long-term radiotoxicity and heat load of high-level waste, thereby easing some of the demands placed on geological repositories~\cite{Salvatores2011PNT,IAEA2025FuelCycleRev2}. Its practical implementation, however, requires accurate knowledge of reaction cross sections, irradiation spectra, fuel-cycle inventories, and material behavior under repeated recycling conditions~\cite{Salvatores2011PNT,Kolos2022NuclearDataNeeds}.

Advanced reactor concepts, including fast reactors and accelerator-driven systems, are often considered key platforms for transmutation strategies. Fast neutron spectra are particularly effective for fissioning heavy actinides, thereby reducing their long-term accumulation. However, these systems introduce additional engineering challenges, including radiation damage to structural materials, complex core physics, and more demanding requirements on fuel fabrication and reprocessing. The optimization of such systems requires a balance among neutron economy, safety margins, fuel cycle efficiency, and materials durability~\cite{Salvatores2011PNT,GenIVFissionReview2023,CIADS,CIADS2,NEA2024SMRWasteWorkshop}. Consequently, transmutation cannot be considered in isolation; it must be embedded within a broader system-level analysis of reactor design and fuel-cycle architecture.

Another critical component of nuclear sustainability is geological disposal of high-level waste. Deep geological repositories are widely regarded as the reference long-term solution for isolating radioactive waste from the biosphere. The scientific basis of geological disposal involves geomechanics, hydrogeology, geochemistry, engineered-barrier performance, and radionuclide transport. Key processes include radionuclide migration through fractured or porous host rocks, chemical interactions between waste forms and geological media, corrosion of engineered barriers, and the long-term evolution of repository environments under thermal and radiological loads~\cite{IAEA2025FuelCycleRev2,NEA2014GeologicalSafetyCase}. These processes must be evaluated over timescales extending far beyond ordinary engineering design horizons, which makes uncertainty quantification and safety-case methodology central to repository assessment~\cite{NEA2014GeologicalSafetyCase}.

Beyond the purely technical aspects, the eighth frontier question explicitly encompasses long-term governance, institutional stability, and societal trust. Unlike most engineering systems, nuclear-waste management requires decision-making frameworks that remain robust across generations. This introduces unique challenges related to regulatory continuity, knowledge preservation, intergenerational responsibility, and institutional resilience. It also raises fundamental questions about how scientific uncertainty, risk perception, and societal values are incorporated into long-term policy decisions~\cite{NEA2022StakeholderConfidence}. In this sense, nuclear-waste management is not only a technical problem, but also a socio-technical system problem that requires sustained interaction among scientists, engineers, regulators, policymakers, and the public.

For this reason, the eighth frontier question should be treated as a first-class research problem rather than as a downstream regulatory afterthought. The integration of science and governance is essential for ensuring that nuclear technologies remain socially acceptable and environmentally responsible over long timescales. Partitioning and transmutation, advanced fuel cycles, and geological disposal each involve tightly coupled interactions among nuclear physics, materials chemistry, reactor engineering, and systems analysis~\cite{Salvatores2011PNT,IAEA2025FuelCycleRev2,NEA2014GeologicalSafetyCase}. At the same time, they depend on institutional frameworks capable of maintaining safety standards, regulatory oversight, knowledge transfer, and public confidence over extended periods~\cite{NEA2022StakeholderConfidence}.

Moreover, the continued expansion of advanced nuclear reactors and isotope-production systems will place increasing demands on fuel cycle infrastructure, waste minimization strategies, and coordinated emergency-response capabilities. As nuclear systems become more diverse and technologically sophisticated, the complexity of associated fuel cycles will also increase, requiring more integrated approaches to resource management and waste handling. This includes not only technical optimization of fuel utilization and recycling pathways, but also strategic planning for supply-chain resilience, material accounting, safeguards, and international cooperation~\cite{IAEA2025FuelCycleRev2,NEA2024SMRWasteWorkshop}. {An operational research agenda for nuclear governance must also include severe-accident source terms, digital safety cases, decommissioning and remote-handling technologies, environmental monitoring networks, emergency-response platforms, cyber-physical resilience, and transparent public communication. These capabilities convert high-level governance principles into testable technical systems.}

In this broader context, a mature and sustainable nuclear future must be evaluated using criteria that extend beyond simple metrics of energy production or isotope output. While these quantities remain important, they are insufficient on their own to characterize the long-term viability of nuclear systems. Equally critical is the ability to manage the full lifecycle of nuclear materials in a technically credible, environmentally sound, and socially durable manner. This includes minimizing long-term environmental impact, ensuring robust safety under both normal and abnormal conditions, and maintaining transparent and trustworthy governance structures.

Ultimately, the challenge of fuel cycle sustainability and nuclear waste management represents a convergence of fundamental science, applied engineering, and long-term societal responsibility \cite{NEA2024SMRWasteWorkshop}. Addressing this challenge requires not only advances in nuclear physics and materials science, but also the development of integrated frameworks that connect technical solutions with institutional design and public policy. In this sense, the eighth frontier question encapsulates one of the most profound dimensions of nuclear science: the need to align powerful technological capabilities with equally robust systems of stewardship across deep time.

\subsection{International cooperation, scientific sovereignty, and strategic positioning}

\begin{figure*}[!htb]
    \includegraphics[width =0.8\linewidth]{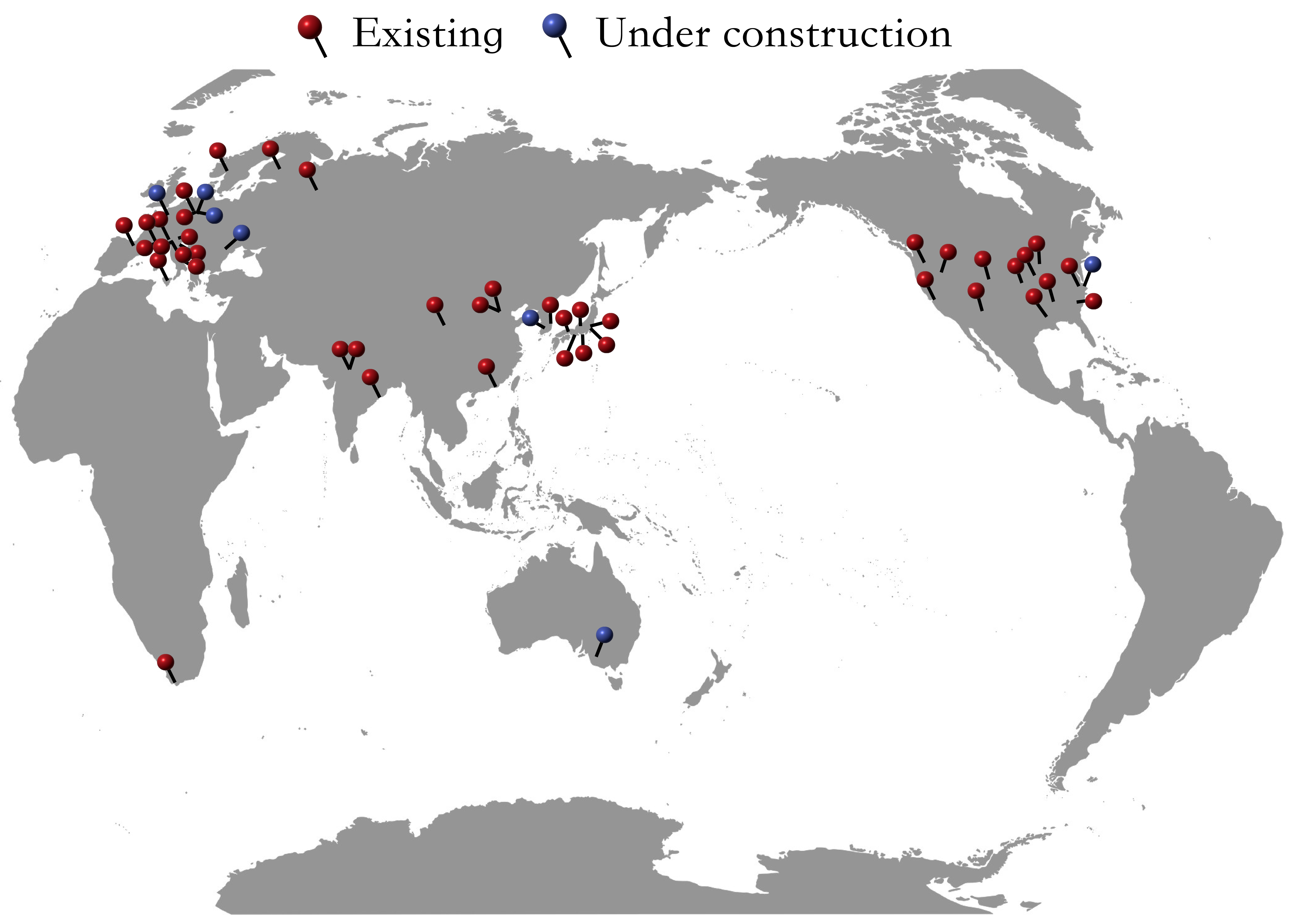}
    \caption{World-wide facilities for nuclear physics. 
{Schematic redrawn and adapted by the author from Ref.~\cite{Ma2026}.}}
    \label{fig:nuclear_facilities}
\end{figure*}

Nuclear science is inherently and irreversibly international in scope. The combination of high-cost infrastructure, large-scale instrumentation, and globally distributed expertise means that most frontier research programs in nuclear physics now depend on multinational collaboration. Large accelerator complexes, next-generation detector arrays, isotope production facilities, and high-performance computing platforms are rarely sustainable within a single institution or even a single nation (see Fig.~\ref{fig:nuclear_facilities}). In addition, the most significant experimental campaigns—such as rare-isotope beam programs, heavy-ion collision experiments, and precision measurements of weak decay processes—typically require coordinated access to beam time, shared detector systems, and globally integrated data evaluation frameworks. As a result, international cooperation is not merely beneficial but structurally necessary for progress in the field.

At the same time, nuclear science occupies a unique position among modern disciplines because of its dual-use nature and its deep connections to national infrastructure, energy systems, medical technologies, and security considerations. Nuclear technologies underpin critical applications ranging from energy generation and medical isotope production to radiation safety and materials testing. Consequently, issues of scientific sovereignty, technological independence, and supply chain resilience play a more prominent role in nuclear science than in many other areas of basic research. This creates a persistent and delicate balance between the open, collaborative nature of scientific discovery and the strategic imperative of maintaining national capability in key technological domains.

The resulting tension between openness and sovereignty is one of the defining features of the current era in nuclear science. On one hand, openness is essential for maximizing scientific return. International collaborations enable the pooling of resources, the sharing of expertise, and the cross-validation of experimental and theoretical results. They also reduce unnecessary duplication of expensive infrastructure and promote standardized methodologies and data formats. On the other hand, strategic autonomy is required to ensure reliable access to critical technologies, maintain continuity of research programs, and safeguard national interests in areas such as energy security, medical isotope supply, and high-end instrumentation development.

In practice, successful long-term strategy in nuclear science is likely to require a carefully structured, multi-layered approach. At the level of fundamental discovery, open international collaboration should remain the dominant paradigm. Large-scale scientific questions—such as the nature of nuclear matter under extreme conditions, the limits of nuclear stability, or the origin of the elements—benefit from globally distributed experimental campaigns and shared theoretical frameworks. International organizations and collaborative experimental facilities play a crucial role in enabling these efforts, ensuring that scientific questions are addressed with maximal statistical power and minimal redundancy.

At the infrastructure level, however, sustained national investment becomes essential. The construction and operation of major accelerator facilities, detector systems, isotope production platforms, and advanced computing resources require long-term financial commitment and strategic planning. Equally important is the development of domestic talent pipelines capable of supporting these infrastructures, including expertise in accelerator physics, detector technology, nuclear theory, data science, and engineering. Without such internal capability, meaningful participation in international collaborations becomes increasingly difficult, and scientific autonomy may be compromised.

At the governance level, nuclear science introduces additional layers of complexity. Issues of nuclear safety, radiation protection standards, waste management protocols, and emergency response systems require stable institutional frameworks and long-term regulatory continuity. Because nuclear technologies can have cross-border environmental and security implications, international communication and coordination are also essential. Effective governance therefore depends on both domestic institutional robustness and active participation in international regulatory and technical bodies. Trust, transparency, and consistency in safety standards are critical components of this framework.

Within this global landscape, countries with rapidly developing nuclear science programs face both opportunities and responsibilities. For China, in particular, the rapid expansion of large-scale scientific infrastructure provides a unique strategic position. The development of advanced facilities such as next-generation accelerator complexes and precision experimental platforms enables deeper participation in global scientific collaborations while simultaneously strengthening domestic research capabilities. HIAF represents not only an experimental tool but also a strategic platform for shaping future directions in rare-isotope physics, nuclear astrophysics, and high-energy-density matter studies \cite{Zhou2022,HIAF,HIAF2,HIAFSHE}. Located in Huizhou and developed by the Institute of Modern Physics under the Chinese Academy of Sciences, it integrates high-intensity stable and radioactive heavy-ion beams, advanced storage rings, superconducting accelerators, and modern detector systems. Its program spans precision studies of exotic nuclei, dense nuclear matter, atomic physics, materials science, and biomedical applications.

As domestic capabilities mature, there is increasing potential not only to contribute to existing international programs, but also to play a more active role in defining scientific agendas in selected subfields. This includes setting priorities for future experimental campaigns, developing novel detector technologies, and contributing to global data evaluation efforts. In some areas, such as exotic nuclei studies, nuclear astrophysics, and heavy-ion collision physics, the combination of large-scale facilities and strong theoretical communities provides a foundation for leadership in specific research directions. At the same time, maintaining strong international engagement remains essential for ensuring scientific excellence and avoiding fragmentation of research efforts.

Another important aspect of strategic positioning is the development of interoperable scientific standards and data-sharing frameworks. As nuclear science becomes increasingly data-intensive and computationally driven, the ability to share experimental results, theoretical models, and evaluated nuclear data across institutional and national boundaries becomes a key factor in scientific efficiency. International collaborations on nuclear data evaluation, for example, play a critical role in ensuring consistency across applications ranging from reactor design to astrophysical modeling. Participation in such frameworks enhances both scientific impact and technological interoperability.

{Recent policy developments in China make this connection between major facilities and international access more concrete. In 2026, the National Natural Science Foundation of China launched a pilot ``International Open Cooperation on National Major Science and Technology Platforms'' program to support foreign researchers working full-time abroad and Chinese researchers in carrying out high-level collaborative basic research in China based on national major science and technology platforms \cite{NSFCMajorPlatform2026}. The call defines eligible platforms as either national major science and technology infrastructure that has been approved, completed, and officially operated by the National Development and Reform Commission (NDRC), or major research instruments developed under NSFC major instrument projects that have passed final acceptance and been integrated into the national network management platform for major research infrastructure and large-scale scientific instruments. Two project categories are offered under the program: key cooperation projects and exploratory cooperation projects, which differ in funding amounts and project durations.}

In addition, the training and mobility of scientific talent constitute a central element of international cooperation. Nuclear science has traditionally relied on a highly mobile and internationally trained workforce, with researchers frequently moving between laboratories, universities, and research institutions across different countries. This mobility facilitates knowledge transfer, fosters collaboration, and helps maintain high scientific standards. Ensuring continued support for such exchanges is therefore essential for sustaining the vitality of the field. 
{The NSFC platform-based pilot program is important not only as an individual funding call but also as an early institutional mechanism for linking major domestic platforms with international user communities \cite{NSFCMajorPlatform2026}. For nuclear science, this provides a policy-level route through which facilities such as HIAF, CJPL/JUNA, SLEGS, CSNS/Back-n, advanced isotope-production platforms, and reactor- or accelerator-based testing infrastructures can host international projects while retaining clear domestic responsibility for safety, data stewardship, and strategic capability. As these platforms mature and user communities expand, more programs that facilitate international joint research, researcher exchange, and shared use of major facilities can be expected to emerge, further strengthening China's role in global nuclear science.}

Ultimately, the long-term evolution of nuclear science will depend on the ability to balance three interdependent dimensions: open scientific collaboration, national strategic capability, and robust international governance. None of these elements can be fully effective in isolation. Open collaboration without sufficient domestic capability risks dependency and fragmentation, while excessive focus on sovereignty may limit scientific progress and international integration. Similarly, governance structures that are not aligned internationally may lead to inefficiencies or inconsistencies in safety and regulatory practices.

In conclusion, international cooperation and scientific sovereignty are not opposing goals, but rather complementary components of a complex global scientific ecosystem. The challenge for the future is to design institutional and strategic frameworks that allow both to be realized simultaneously. In this context, countries with strong and rapidly developing nuclear science infrastructures are positioned not only as participants in global research efforts, but also as potential contributors to the shaping of the international scientific landscape itself. Such a role carries both opportunities and responsibilities, reinforcing the importance of long-term strategic planning, sustained investment, and active engagement in global scientific governance.

\subsection{{Recent Chinese progress in nuclear technologies}}

{The examples distributed across the preceding sections can be consolidated into a technology chain extending from commercial deployment to research infrastructure. In nuclear energy, Hualong One demonstrates Generation-III engineering and localization at fleet scale; HTR-PM establishes a modular high-temperature gas-cooled route; ACP100 tests multipurpose small-reactor deployment; and CFR-600, TMSR, and CiADS pursue fast-spectrum fuel-cycle closure, thorium chemistry, and accelerator-driven transmutation at demonstration or experimental stages. EAST's 1066-s high-confinement result and the CRAFT component-test program similarly connect plasma performance to fusion nuclear engineering. These platforms should be assessed not as isolated devices but through common needs in nuclear data, materials qualification, chemistry, tritium and source-term control, licensing, and digital instrumentation.}

{In medicine and isotope science, the scale of China's clinical nuclear-medicine community creates a direct demand for reliable $^{99}$Mo, $^{177}$Lu, $^{64}$Cu, $^{89}$Zr, $^{225}$Ac, and other diagnostic and therapeutic radionuclides. Reactor and accelerator production must be coupled to enriched targets, radiochemical separation, metrology, quality assurance, logistics, and clinical trials. Domestic heavy-ion therapy provides a complementary example of translation from accelerator physics and dosimetry to certified medical equipment and hospital operation.}

{Nuclear-data and irradiation infrastructure form the shared middle layer of this chain. SLEGS and CSNS/Back-n support photonuclear and neutron-induced measurements, but their lasting value depends on evaluated libraries, covariance data, benchmark experiments, detector-response models, and reproducible open workflows. The proposed Tsinghua High Flux Reactor (THFR), designed for thermal and fast neutron fluxes of approximately $2\times10^{15}$~n\,cm$^{-2}$\,s$^{-1}$, would add fuel and structural-material irradiation, high-specific-activity isotope production, and neutron-science capability~\cite{THFR2025}.}

{Finally, deployment at scale requires safety and governance technologies to advance with hardware. Severe-accident source-term modeling, digital safety cases and reactor twins, predictive maintenance, decommissioning, environmental monitoring, emergency response, cybersecurity, and public communication should be developed as interoperable capabilities. This cross-cutting view also clarifies the role of AI: it can connect autonomous experiments, nuclear-data evaluation, dose optimization, equipment health, and safety monitoring, provided that physical constraints, traceable data, and uncertainty quantification remain explicit.}

\section{Outlook}

The ten frontier questions in nuclear science and technology provide a coherent roadmap precisely because they resist disciplinary fragmentation and instead emphasize deep structural unity across the field. Rather than separating nuclear structure, reactions, astrophysics, high-energy nuclear matter, energy applications, and enabling technologies into independent subdisciplines, the framework highlights their intrinsic interdependence (see Fig.~\ref{fig:frontier_nuclear_question}). It reminds us that nuclear science is ultimately governed by a small number of foundational challenges: how the strong interaction gives rise to emergent structure across vastly different length and energy scales; how finite quantum many-body systems reorganize themselves near thresholds, at limits of stability, and under extreme thermodynamic or dynamical conditions; how astrophysical observations and terrestrial experiments constrain and inform one another; how instrumentation, computation, and theory co-evolve as a unified technological system; and how a fundamental physical science can simultaneously underpin long-term societal needs in energy, medicine, security, and materials.

Looking toward the next decade, several interrelated directions are likely to play a particularly consequential role in shaping the evolution of the field. First, the traditional boundaries between nuclear structure, reaction theory, and continuum dynamics will continue to weaken. In rare-isotope science, weak binding and proximity to decay thresholds naturally demand descriptions that unify bound states, resonances, and scattering channels within a single theoretical framework. Similarly, in nuclear astrophysics, reaction networks are increasingly sensitive to properties of nuclei far from stability, where structure and reaction mechanisms cannot be cleanly separated. This gradual unification of previously distinct conceptual domains is likely to redefine how nuclear phenomena are categorized and understood.

Second, precision will become an even more dominant organizing principle across the entire field. Whether in the determination of nuclear masses near the driplines, the measurement of fluctuation observables in relativistic heavy-ion collisions, the characterization of nuclear electromagnetic moments, or the development of nuclear clocks and radiometric standards, future progress will depend on pushing uncertainties to systematically lower levels. Importantly, such precision cannot be achieved through instrumentation alone. It will increasingly require tightly integrated experimental platforms in which accelerators, detectors, data acquisition systems, and theoretical models are co-designed to minimize systematic uncertainties and maximize information extraction. In this sense, precision nuclear science is evolving toward a systems-engineering paradigm rather than a purely instrument-driven one.

Third, data-driven and AI-assisted methodologies will become standard components of both experimental and theoretical workflows. However, their long-term scientific value will depend critically on their integration with physical constraints and rigorous uncertainty quantification. Purely phenomenological machine learning approaches will be insufficient in a domain where extrapolation beyond available data is often essential. Instead, the most impactful developments are likely to emerge from hybrid frameworks that embed symmetries, conservation laws, and effective field theory structures into data-driven models. In parallel, interpretable AI and uncertainty-aware emulators will play a central role in bridging computationally intensive nuclear many-body calculations with large-scale applications in astrophysics, reactor physics, and reaction theory.

Fourth, application-oriented areas such as advanced nuclear energy systems, isotope production technologies, nuclear medicine, and waste management will increasingly be recognized as intellectually rich research frontiers in their own right, rather than purely downstream engineering domains. As discussed throughout this framework, these areas are deeply intertwined with fundamental nuclear physics through shared dependencies on nuclear data, many-body theory, transport modeling, and materials under extreme conditions. Treating them as scientific frontiers ensures that advances in applications feed back into fundamental understanding, while also aligning basic research with long-term societal needs.

From a broader perspective, the most important opportunity created by the ten-question framework is not limited to improved organization of scientific topics or more effective communication across subfields. Its deeper significance lies in enabling a coordinated vision for the future of nuclear science, in which fundamental discovery, experimental facility development, theoretical innovation, and strategic application are treated as mutually reinforcing components of a single evolving ecosystem. Such an integrated perspective makes it possible to align decisions about major infrastructure investment, methodological development, and scientific prioritization in a consistent long-term strategy.

Ultimately, sustaining the vitality and relevance of nuclear science will depend on maintaining this balance between curiosity-driven exploration and application-oriented development, between specialization and integration, and between national initiatives and global collaboration. If successfully achieved, the coming decade may witness not only significant advances in specific subfields, but also a deeper unification of nuclear science as a whole, both conceptually and institutionally. In this sense, the ten frontier questions do not merely describe where the field is going; they help define what it means for nuclear science to progress in a coherent and sustainable way.

\section*{Acknowledgments} This perspective and review for frontier ten questions and emerging directions in nuclear science and technology is under the chairmanship of Prof. Yu-Gang Ma, the editorial boards of ``Nuclear Science and Techniques'' and ``Nuclear Technology'' solicited proposals from leading editorial teams both domestically and internationally, conducted extensive deliberations, and, following approval by the Executive Council of the Chinese Society of Nuclear Physics, finalized the ``Ten Key Questions on the Frontiers of Nuclear Science and Technology.'' Author acknowledges the contribution from above-mentioned team. In particular, author appreciates Dr. Simin Wang for organization  of some figures and texts. 
This work was partially supported by the National Natural Science Foundation of China under Contract  No.\,12547102.

\end{CJK*}
\bibliography{nst_frontier_review_refs}
\end{document}